\def\BE{\begin{equation}}
\def\EE#1{\label{#1}\end{equation}}
\def\be{\begin{align}}
\def\ba{\begin{align*}}
\def\se#1{\begin{subequations}\label{#1}
\renewcommand{\theequation}{\theparentequation.\arabic{equation}}}
\def\rf#1{(\ref{#1})}
\def\xf#1{Fig.~\ref{#1}}
\def\R{{\mathbb R}}
\def\T{{\mathbb T}}
\def\I{{\rm i}}
\def\e{{\rm e}}
\def\d{{\rm d}}
\def\L{\boldsymbol{\mathfrak{L}}}
\def\Ae{\mathfrak{A}}
\def\A{\boldsymbol{\mathfrak{A}}}
\def\sAe{{^{\rm s}\mathfrak{A}}}
\def\sA{{^{\rm s}\boldsymbol{\mathfrak{A}}}}
\def\De{\mathfrak{D}}
\def\D{\boldsymbol{\mathfrak{D}}}
\def\sD{{^{\rm s}\!D}}
\def\mD{{^{\rm s}\bf D}}
\def\M{\boldsymbol{\mathfrak{M}}}
\def\LA{\left\langle}
\def\RA{\right\rangle}
\def\ls{\apprle}
\documentclass[12pt,a4paper]{Melsarticle}
\usepackage{amsmath,amssymb,graphicx,wasysym}
\begin{document}
\title{Magnetic field generation by pointwise zero-helicity
three-dimensional steady flow of incompressible electrically conducting fluid}
\author[mitp]{A.~Rasskazov}
\author[por,sam]{R.~Chertovskih}
\author[mitp]{V.~Zheligovsky}
\address[mitp]{Institute of Earthquake Prediction Theory and
Mathematical Geophysics, Russian Ac. Sci.,\\
84/32 Profsoyuznaya St, 117997 Moscow, Russian Federation}
\address[por]{Research Center for Systems and Technologies, Faculty of
Engineering, University of Porto,\\
Rua Dr.~Roberto Frias, s/n, 4200-465, Porto, Portugal}
\address[sam]{Samara National Research University, 34 Moskovskoye Ave., Samara,
Russian Federation}

\begin{abstract}
We introduce six families of three-dimensional space-periodic steady solenoidal
flows, whose kinetic helicity density is zero at any point. Four families are
analytically defined. Flows in four families have zero helicity spectrum.
Sample flows from five families are used to demonstrate numerically that
neither zero kinetic helicity density, nor zero helicity spectrum prohibit
generation of large-scale magnetic field by the two most prominent dynamo
mechanisms: the magnetic $\alpha$-effect and negative eddy diffusivity.
Our computations also attest that such flows often generate small-scale field
for sufficiently small magnetic molecular diffusivity. These findings indicate
that kinetic helicity and helicity spectrum are not the quantities controlling
the dynamo properties of a flow regardless of whether scale separation
is present or not.

\end{abstract}

\maketitle
\section{Introduction}\label{intr}

Consider a volume $\Omega$ transported by an ideal incompressible fluid flow
$\bf v$, such that initially the vorticity $\nabla\times\bf v$ is tangent
to the boundary $\partial\Omega$. Since the vorticity lines are frozen
in the fluid, the vorticity remains tangent to $\partial\Omega$ at all
times, and the value
$$\int_\Omega{\bf v}\cdot(\nabla\times{\bf v})\,\d{\bf x},$$
called kinetic helicity, does not change in time. This was shown
for barotropic gas in \cite{Mor} and independently (see \cite{Mof} and
https://sites.google.com/site/hkeithmoffatt/selected-publications-1960s)
in \cite{M69} (see also \cite{KhCh,MT,AK}).
In space-periodic flows, the total kinetic helicity in a periodicity
cell is also conserved. Recent measurements of the total helicity of vortex
tubes in water demonstrated that even in viscous fluid the total helicity can
remain constant or saturate to a constant (or almost constant) value \cite{SRK}.

In a general setup, helicity $\cal H$ of a solenoidal vector field
transported by a flow as a frozen field (e.g., a magnetic field or the flow
vorticity) is defined as the volume-integrated scalar product of the field
and its vector potential (by this definition, the kinetic helicity is
the vorticity helicity). Following \cite{MR}, consider a tube carrying flux
$\Phi$ and consisting of closed field lines that have no inflexion points (i.e.,
whose curvature does not vanish at any point). It can be proven that for such
a tube
\BE{\cal H}=n\Phi^2,\EE{WTN}
where $n$ is a conserved quantity. This invariant
quantifies a fundamental topological property of the field, the knottedness
of its field lines. If the tube under consideration is unknotted, but any pair
of field lines in it is linked, the linking number, $\cal N$, being the same
for each pair, then $n=\cal N$. In general, $n={\cal W}+{\cal T}+{\cal N}$,
where the writhe, $\cal W$, and the normalised total torsion of a center
field-line, $\cal T$, characterise how the tube itself is knotted (see
an intuitive illustration of these concepts in Figs.~7 and 12 of \cite{MR}).

Containing important information about the topological structure of the flow,
helicity plays a significant role in physics (see the references
in \cite{KFDI}). In geophysical or astrophysical contexts, as well as in
laboratory experiments, flows are often accompanied by rotation and
consequently they are usually helical. Helicity arises in convection
\cite{LB,Ge} and turbulence \cite{PM}. Participation of helicity in generation
of a dipolar magnetic field in a geodynamo model involving convection
in a rapidly rotating spherical shell was explored in \cite{SJ}. Generation
of cosmic magnetic fields of the intensity and spatial scale that is observed
in astrophysics is believed to be possible due to the chirality of the
background turbulence characterised by a non-zero kinetic helicity \cite{Mof}.
Moreover, apparently the dynamo action of a flow is guaranteed if the mean
helicity is of constant sign over a sufficiently large extent of fluid
\cite{Mof}.

Investigating the electromotive force (e.m.f.) due to the interaction
of small-scale fluctuations of the flow and magnetic field is a pillar of the
magnetic dynamo theory. The small-scale e.m.f.~may have a non-zero mean
component parallel to the mean magnetic field, and this is often beneficial
for magnetic field generation. This seminal idea goes back to E.~Parker
\cite{P55}, who called such fluctuations of the flow ``cyclonic events''.
The part of the mean e.m.f.~linear in the mean field gives rise
to the so-called magnetic $\alpha$-effect. A systematic treatment of this idea
under various simplifying assumptions is the topic of mean-field
electrodynamics \cite{SKR,KR}.

Clearly, an intricate spatial structure of the small-scale fluctuating
component of a flow (and hence, by virtue of the induction equation,
of the magnetic field) is expected to correlate with a high degree
of the vorticity knottedness and --- since this hydrodynamic invariant
constrains the topology of vorticity lines ---
with a non-zero kinetic helicity. Thus, kinetic helicity may be intimately
related with magnetic field generation --- for instance, it may control
the strength of the magnetic $\alpha$-effect. Some observations confirm this
conjecture: On the one hand, the $\alpha$-effect coefficient was calculated
in the second-order correlation approximation and high-conductivity limit
for the isotropic turbulence (see, e.g., \cite{KHR}), and it turned out to be
proportional to the mean kinetic helicity. This result was also derived
in \cite{M74} in the limit of ideal magnetohydrodynamics for rotationally
symmetric turbulence with the use of Lagrangian coordinates for description
of the evolution of the field. On the other, by the Zeldovich
\cite{Zel} antidynamo theorem, a two-dimensional flow of incompressible fluid
cannot generate magnetic field --- and its mean kinetic helicity is zero
if the flow is space-periodic or satisfies some other suitable boundary
conditions; moreover, it is pointwise non-helical,
\BE{\bf v}\cdot(\nabla\times{\bf v})=0,\EE{non}
when it does not depend on the coordinate in the direction perpendicular to
the parallel planes, to which the fluid motion is confined.

However, it became clear decades ago that the mean helicity is unnecessary
for the dynamo action of smooth (laminar) flows, see, e.g., \cite{GFP,RBr}. More
specifically, a non-zero kinetic helicity is required neither for generation
of small-scale (i.e., having the same spatial periods as the flow velocity)
magnetic field, nor for creating the $\alpha$-effect for generation
of the large-scale magnetic field. Not much helicity is needed to drive
large-scale nonlinear dynamos \cite{GBMP}. Nevertheless, parallels
between the generation in various magnetohydrodynamic
(MHD) setups and a non-zero kinetic helicity
of the generating flow are often drawn in the literature. The following argument
is often encountered: the $\alpha$-effect requires the lack of reflectional
symmetry in the flow, and for physicist kinetic helicity is the simplest
measure of this (since the helicity vanishes for parity-invariant and
mirror-symmetric flows). This point of view is amenable to the following
criticism: first, many other functionals, such as the flow helicity (or
the helicity of any real power of the Laplacian of the flow or vorticity)
have the same property; and, second, the total kinetic helicity also vanishes
for flows that are parity-antiinvariant or have reflectional antisymmetry.
Similarly, helicity spectrum vanishes not only for parity-invariant flows,
but also for parity-antiinvariant ones (see section \ref{hsf}).

The discussion above implies that exploring the dynamics of ideal fluid flow or
evolution of magnetic field (note that formally vorticity satisfies the same
equation as the magnetic field) transported by flows of complex topology
may require constructing flows with a desirable knot structure of field lines.
A systematic procedure for constructing solenoidal vector fields with tunable
helicity, whose closed field lines involve knots of many types, was presented
in \cite{KFDI}. It relies on a method for calculating vector potentials
of the fields that employs complex scalar functions. Examples
of knotted fields were presented, whose mean helicity is zero.

We focus on the study of the kinematic dynamo action of steady three-dimensional
flows of incompressible fluid and demand that the flow ultimately lacks kinetic
helicity, the flow and vorticity being orthogonal at each point \rf{non};
we will call such flows non-helical. In section~\ref{const} we present
six families of steady solenoidal pointwise non-helical flows. By necessity,
our approach is constructive: flows from four families are analytically defined,
from another one can be obtained by semianalytical procedures. Sample flows
belonging to each of the five families are used to study kinematic dynamos.
Four families out of the five are composed of flows, whose helicity spectrum
is zero; we therefore simultaneously verify that a non-zero helicity spectrum
is unnecessary for generation of large- or small-scale flows.

Let us mention the small-scale kinematic dynamo \cite{ZG}
powered by the Christopherson flow \cite{Chr}:
\BE{\bf v}=\left({L^2\over4\pi}{\partial g\over\partial x_1}\cos\pi x_3,\quad
{L^2\over4\pi}{\partial g\over\partial x_2}\cos\pi x_3,\quad
{g\over3}\sin\pi x_3\right),\EE{Chris}
where
$$g(x_1,x_2)=\cos{2\pi x_1\over\sqrt{3}L}+2\cos{\pi x_1\over\sqrt{3}L}\cos{\pi x_2\over L}.$$
It satisfies \rf{non} and nevertheless generates magnetic field when
the magnetic Reynolds number exceeds the critical value $R_m\approx515.63$
\cite{ZG}. Matthews \cite{Mat} questioned whether the resolution in computations
\cite{ZG} was sufficient. We have now repeated the computations with the double
resolution of $128^3$ poloidal and $128^3$ toroidal modes, and recovered
the magnetic field growth rates of \cite{ZG}, increasing from -0.000377 for
$R_m=500$ to 0.006786 for $R_m=1000$, with an accuracy better than $10^{-6}$,
this confirming the results of \cite{ZG}. The energy spectrum of the dominant
magnetic eigenmodes falls off by at least 11 orders of magnitude
indicating that the resolution that we have now used is exceedingly high.
The estimate $R_m=515.63$ for the critical value was obtained in \cite{ZG}
by linear interpolation of growth rates between the two neighbour integer
magnetic Reynolds numbers; the present double resolution computations
yield for this $R_m$ the growth rate $1.2\times10^{-5}$.

Many examples of dynamos can be found in the literature, in which the total kinetic
helicity vanishes. However, to the best of our knowledge,
the Christopherson flow is the only documented example of a three-dimensional
steady flow of incompressible fluid, whose helicity density is zero at each point
in space \rf{non}, and which is capable of dynamo action. This example is
unsatisfactory in that it is a dynamo only for specific boundary conditions
(see \cite{Mat}). We consider only space-periodic flows and magnetic fields
in order to exclude the influence of boundaries, the flow periodicity cell
being a cube $\T^3=[0,2\pi]^3$. Our goal is to demonstrate that
it is typical for a pointwise non-helical steady flow to generate magnetic field
for a sufficiently small magnetic molecular diffusivity, regardless of whether
scale separation is present or not. In particular, we will show that such flows
typically can power the two most prominent mechanisms for generation
of large-scale fields: the magnetic $\alpha$-effect when the flow lacks parity
invariance, or negative magnetic eddy diffusivity otherwise. It turns out that
many of these flows can also act as small-scale kinematic dynamos.

The paper is organised as follows.
We present large-scale kinematic dynamos based on the magnetic $\alpha$-effect
in section \ref{al}, and negative magnetic eddy diffusivity in section \ref{ed}.
We need a sufficient stock of non-helical flows for numerical experimentation,
and in section \ref{const} we discuss semianalytical techniques
for constructing them, as well as present analytical examples of such flows.
In a multiscale setup, the magnetic $\alpha$-effect and eddy diffusivity tensors
have been calculated by asymptotic methods (see, e.g., chapter 3 of \cite{VZ}).
Our approach relies on this analysis and, for the reader's convenience, in the next
section we summarise and enhance it tailoring to the needs of the present
investigation. The notion of the helicity spectrum arises naturally
in the theory of the magnetic $\alpha$-effect in MHD turbulence \cite{M70,MP};
we briefly discuss in section \ref{hs} its relevance when the local magnetic
Reynolds number is small and evaluate in section \ref{hsf} the helicity
spectrum of the flows that we use in computations.

\section{The magnetic $\alpha$-effect and eddy diffusivity}\label{mult}

We review here the multiscale formalism arising in the study of the kinematic
generation of large-scale magnetic field by a small-scale zero-mean
steady flow $\bf v$ of conducting fluid (see \cite{VZ} for a more detailed
discussion and a comprehensive list of references). In this section no
assumptions about the kinetic helicity are made.

In mathematical terms, we consider the eigenvalue problem for the magnetic
induction operator
\BE\L{\bf b}\equiv\eta\nabla^2{\bf b}+\nabla\times({\bf v}\times{\bf b})=\lambda{\bf b}.\EE{Leig}
Here $\eta$ is the magnetic molecular diffusivity, $\bf b$ a magnetic mode
and Re\,$\lambda$ its growth rate (a negative growth rate actually indicates
that a mode is decaying). The mode is solenoidal,
\BE\nabla\cdot{\bf b}=0,\EE{bs}
and the fluid is supposed to be incompressible, $\nabla\cdot{\bf v}=0$.

The two-scale nature of the magnetic mode $\bf b$ is reflected by its
dependence on the so-called fast, $\bf x$, and slow, ${\bf X}=\varepsilon\bf x$,
spatial variables; by contrast, the small-scale flow $\bf v$
depends only on~$\bf x$. The scale ratio $\varepsilon$ is assumed to be small,
which enables us to apply asymptotic
methods. (The difference in approaches is notable: while we consider the limit
$\varepsilon\to0$ only, the theory of mean-field electrodynamics strives
to estimate the impact of all small and intermediate scales on the
large-scale magnetic field, somewhat in the spirit of the LES closures;
see \cite{ABNZ}.) One proceeds by substituting power series expansions
\se{ble}\be{\bf b}&=\sum_{n=0}^\infty{\bf b}_n({\bf X},{\bf x})\,\varepsilon^n,\label{bex}\\
\lambda&=\sum_{n=0}^\infty\lambda_n\varepsilon^n\label{lex}\end{align}\end{subequations}
into \rf{Leig} and \rf{bs}, and deriving a hierarchy of equations emerging at
successive orders $\varepsilon^n$.

\subsection{Magnetic $\alpha$-effect}\label{mae}

The relevant solution to the first (order $\varepsilon^0$) equation
in the hierarchy is
$${\bf b}_0=\sum_{k=1}^3\LA{\bf b}_0\RA_k({\bf e}_k+{\bf S}_k),\qquad\lambda_0=0,$$
where
$$\LA{\bf f}\RA=(2\pi)^{-3}\int_{\T^3}{\bf f}({\bf X},{\bf x})\,\d{\bf x}
=\sum_{k=1}^3\LA{\bf f}\RA_k{\bf e}_k$$
denotes the mean over the periodicity cell $\T^3$ in the fast variables (i.e.,
over small scales; note that no other means are appropriate), ${\bf e}_k$
are unit vectors of the Cartesian coordinate system and ${\bf S}_k({\bf x})$
are zero-mean small-scale solenoidal solutions to 3 {\it auxiliary problems
of type I}:
\BE\L{\bf S}_k=-{\partial{\bf v}\over\partial x_k}\qquad
\Leftrightarrow\qquad\L({\bf S}_k+{\bf e}_k)=0\EE{Seq}
(the magnetic induction operator $\L$ is henceforth assumed to involve
differentiation in fast variables $\bf x$ only; we will call it
the large-scale magnetic induction operator when full differentiation in space
is performed). It is easy to show that such
solutions exist, see, e.g., section 3.2 of \cite{VZ}.

The solvability condition for the second (order $\varepsilon^1$) equation
in the hierarchy is an eigenvalue problem,
\BE\nabla_{\bf X}\times(\A\LA{\bf b}_0\RA)=\lambda_1\LA{\bf b}_0\RA,\qquad
\nabla_{\bf X}\cdot\LA{\bf b}_0\RA=0,\EE{aleq}
(the subscript $\bf X$ marks differential operators in slow variables)
from which we determine $\lambda_1$ and, generically, $\LA{\bf b}_0\RA$.
Here $\A$ is the so-called tensor of the magnetic $\alpha$-effect, a $3\times3$
matrix, whose $k$th column is
\BE\A_k=\LA{\bf v}\times{\bf S}_k\RA.\EE{Adef}
This expression is consistent with the original Parker's idea that
the interaction of fine structures of a flow (in our case, $\bf v$) and magnetic
field ($\sum_{k=1}^3\LA{\bf b}_0\RA_k{\bf S}_k$) gives rise to a mean
e.m.f.~($\A\LA{\bf b}_0\RA$) that may have a component, parallel
to the large-scale magnetic field ($\LA{\bf b}_0\RA$, respectively).

To solve the eigenvalue problem \rf{aleq}, we assume that the mean field is
a Fourier harmonics,
\BE\LA{\bf b}_0\RA={\bf B}\e^{\I\bf q\cdot X}.\EE{mm}
Here $\bf q$ and $\bf B$ are constant vectors, $|{\bf q}|=1$. It is convenient
to express the wave vector in spherical coordinates, whose axis is aligned
with the Cartesian axis $x_3$ (the assumed vertical direction):
\se{Bq}\BE q_1=\sin\theta\,\cos\varphi,\qquad q_2=\sin\theta\,\sin\varphi,\qquad
q_3=\cos\theta.\EE{wv}
Since the solenoidality of $\LA{\bf b}_0\RA$ translates into
the orthogonality ${\bf B\cdot q}=0$, we can expand
\BE{\bf B}=B_t{\bf q}^{\rm t}+B_p{\bf q}^{\rm p},\EE{exp}
where unit vectors
\BE{\bf q}^{\rm p}=(\cos\theta\,\cos\varphi,\,\cos\theta\,\sin\varphi,\,-\sin\theta),
\qquad{\bf q}^{\rm t}=(-\sin\varphi,\cos\varphi,0)\EE{uv}\end{subequations}
constitute, together with $\bf q$, an orthonormal basis of positive orientation
in $\R^3$. This reduces \rf{aleq} to an eigenvalue problem for a $2\times2$ matrix:
\BE\I\left[\begin{array}{rr}
{\bf q}^{\rm p}\cdot\A{\bf q}^{\rm t}&{\bf q}^{\rm p}\cdot\A{\bf q}^{\rm p}\\
-{\bf q}^{\rm t}\cdot\A{\bf q}^{\rm t}&-{\bf q}^{\rm t}\cdot\A{\bf q}^{\rm p}\end{array}\right]
\left[\begin{array}{l}B_t\\B_p\end{array}\right]
=\lambda_1({\bf q})\left[\begin{array}{l}B_t\\B_p\end{array}\right].\EE{Am}
The eigenvalues are now obtained by straightforward algebra (the identity
$q^{\rm p}_nq^{\rm t}_{n'}-q^{\rm t}_nq^{\rm p}_{n'}=\epsilon_{jnn'}q_j$
is useful, where $\epsilon_{jln}$ is the unit antisymmetric tensor).
In terms of wave vector components they are
\se{ei}\be\lambda_{1_\pm}({\bf q})=&\,{\I\over2}\left((\Ae^2_3-\Ae^3_2)q_1
+(\Ae^3_1-\Ae^1_3)q_2+(\Ae^1_2-\Ae^2_1)q_3\vphantom{^2}\right)\pm\sqrt a,\label{lpm}\\
a=&\,(\Ae_2^2\Ae_3^3-(\sAe^2_3)^2)q_1^2+(\Ae_1^1\Ae_3^3-(\sAe^1_3)^2)q_2^2
+(\Ae_1^1\Ae_2^2-(\sAe^1_2)^2)q_3^2\label{D}\\
&+2\left((\sAe^1_2\sAe^1_3-\sAe^2_3\Ae_1^1)q_2q_3\vphantom{^2}
+(\sAe^1_2\sAe^2_3-\sAe^1_3\Ae_2^2)q_1q_3
+(\sAe^1_3\sAe^2_3-\sAe^1_2\Ae_3^3)q_1q_2\right),
\nonumber\end{align}\end{subequations}
where $\sAe_i^j=(\Ae_i^j+\Ae_j^i)/2$ are entries of the symmetrised
$\alpha$-tensor $\sA=(\A+\A^*)/2$. Comparison of \rf{D} with the formula
for $\sA^{-1}$ (provided $\sA$ is invertible) in terms
of cofactors (see, e.g., \cite{Shi}) reveals a compact expression
$$a={\bf q}\cdot(\det\sA)\,\sA^{-1}\bf q.$$

For $a\le0$, the $\alpha$-effect just sustains constant-amplitude oscillations
of a mean magnetic mode \rf{mm} in the slow time $T_1=\varepsilon t$. Consider
now wave vectors $\bf q$ for which $a>0$. By virtue of \rf{lpm}, the slow-time
growth rate Re$\,\lambda_{1_+}({\bf q})=\sqrt a$ of the large-scale magnetic
field depends only on the entries of the symmetric matrix $\sA$. This helps
to determine the maximum growth rate
Re$\,\lambda_1({\bf q})$ over unit wave vectors: Eigenvalues $\alpha_i$
of $\sA$ are real and the associated eigenvectors are mutually
orthogonal. The relation \rf{D} remains applicable in the Cartesian coordinate
system, whose axes coincide with eigendirections of $\sA$; thus
$$a=\alpha_1\alpha_2(q'_3)^2+\alpha_2\alpha_3(q'_1)^2
+\alpha_1\alpha_3(q'_2)^2,$$
where $q'_i$ are components of $\bf q$ in the basis of the eigenvectors of $\sA$
(cf.~section 9.3 of \cite{Mb}). Denoting the maximum slow-time growth rate
due to the action of the $\alpha$-effect by $\gamma_\alpha$, we find
\BE\gamma_\alpha\equiv\max_{|{\bf q}|=1}{\rm Re}\,\lambda_{1_\pm}({\bf q})=
\sqrt{\max(\alpha_1\alpha_2,\,\alpha_2\alpha_3,\,\alpha_1\alpha_3)}.\EE{mgr}

A few comments stemming from \rf{ei} are in order.

$i$. The spectrum of the $\alpha$-effect operator,
${\bf b(X)}\mapsto\nabla_{\bf X}\times(\A{\bf b})$, is symmetric about
the real and imaginary axes. Generically, $\A$ is a non-symmetric matrix; thus,
the temporal growth or decay of a mean magnetic mode is accompanied
by oscillations in the slow time $T_1$, whose frequency is controlled
by the antisymmetric part, $\A-\sA$, of the $\alpha$-tensor.

$ii$. The maximum slow-time growth rate \rf{mgr} is strictly positive unless
an eigenvalue of $\sA$ is zero, another one is non-negative, and the third one
is non-positive. If all the three $\alpha_i$ have the same signs, then for each
$\bf q$ there exist a growing and decaying mean magnetic mode. If two
eigenvalues have the same sign and the third one has the opposite sign, then
for some $\bf q$ both modes experience constant-amplitude oscillations
in the slow time $T_1$. The wave vectors for which the modes exhibit such
a purely oscillatory behaviour form a cone, whose cross-section is
elliptic and whose axis is aligned with the eigenvector associated with
the third eigenvalue.

$iii$. When $\A$ is the identity matrix, $\lambda_{1_\pm}({\bf q})=\pm1$
for all unit wave vectors. In particular, the proper subspace associated
with the eigenvalue 1 for ${\bf q}=\pm{\bf e}_i$ is six-fold. ABC-flows
\cite{Arn} and their spatial derivatives constitute a basis in it.

Provided the eigenvalues \rf{ei} are distinct (i.e., $a\ne0$) and the magnetic
induction operator $\L$ does not have zero-mean neutral modes (generically
the two conditions hold true), all terms in the series \rf{ble} can be determined
from the hierarchy of equations obtained by substituting the series into
\rf{Leig} and \rf{bs}. It was proven in \cite{V87} that if the symmetrised
tensor $\sA$ is positively or negatively defined (and if the spatial
periodicity of the eigenfunction is compatible with that of the flow), then the
series \rf{ble}, constructed for any solution \rf{mm}, \rf{ei} to the eigenvalue
problem \rf{aleq} for the $\alpha$-effect operator, converge for sufficiently
small $\varepsilon$ in suitable Sobolev spaces; the sums are analytical
in $\varepsilon$ functions that solve the eigenvalue problem
for the large-scale magnetic induction operator. In other words, a unique
$\varepsilon$-parameterised branch of the eigenvalues \rf{lex} of the large-scale
magnetic induction operator originates from any simple eigenvalue $\lambda_1$
of the $\alpha$-effect operator.

``Uncurling'' \rf{Seq}, we find
\BE-\eta\,\nabla\times{\bf S}_k+{\bf v}\times({\bf S}_k+{\bf e}_k)=
\LA{\bf v}\times{\bf S}_k\RA+\nabla p_k,\EE{unc}
where $p_k({\bf x})$ are space-periodic functions. This identity was used
in \cite{VZ} (see p.~34) to demonstrate that the $\alpha$-effect is linked
with the helicity of the current density $\nabla\times{\bf S}_k$: scalar
multiplying \rf{unc} by ${\bf S}_l+{\bf e}_l$ and averaging the result
over the periodicity cell yields
$$-\eta\,\LA{\bf S}_l\cdot\nabla\times{\bf S}_k\RA
+\LA({\bf S}_l+{\bf e}_l)\cdot({\bf v}\times({\bf S}_k+{\bf e}_k))\RA=\Ae^l_k,$$
whereby
\se{cS}\BE-2\eta\,\LA{\bf S}_l\cdot\nabla\times{\bf S}_k\RA=\Ae^l_k+\Ae^k_l;\EE{cSh}
for $k=l$ this relation reduces to
\BE-\eta\,\LA{\bf S}_k\cdot\nabla\times{\bf S}_k\RA=\Ae_k^k.\EE{cSd}\end{subequations}

The identity \rf{unc} also implies a relation between the tensors
$\A$ and $\A^-$ of the magnetic $\alpha$-effect for a flow $\bf v$ and
the reverse flow $-\bf v$, respectively. Let us denote by $\L^-$
the magnetic induction operator for the reverse flow $-\bf v$,
\BE\L^-{\bf b}=\eta\nabla^2{\bf b}-\nabla\times({\bf v}\times{\bf b})\EE{Lm}
(involving differentiation in the fast variables $\bf x$ only),
whose kernel is spanned by the neutral modes ${\bf S}^-_l+{\bf e}_l$:
\BE\L^-({\bf S}^-_l+{\bf e}_l)=0,\qquad\LA{\bf S}^-_l\RA=0.\EE{Sml}
This equation is equivalent to
\BE-\eta\,\nabla\times{\bf S}^-_l-{\bf v}\times({\bf S}^-_l+{\bf e}_l)=
\LA-{\bf v}\times{\bf S}^-_l\RA+\nabla p^-_l.\EE{cnu}
Scalar multiplying \rf{unc} by ${\bf S}^-_l+{\bf e}_l$
and averaging the result over the periodicity cell yields
$$-\eta\,\LA{\bf S}^-_l\cdot\nabla\times{\bf S}_k\RA
+\LA({\bf S}^-_l+{\bf e}_l)\cdot({\bf v}\times({\bf S}_k+{\bf e}_k))\RA=\Ae^l_k.$$
Similarly, from \rf{cnu}
$$-\eta\,\LA{\bf S}_k\cdot\nabla\times{\bf S}^-_l\RA
-\LA({\bf S}_k+{\bf e}_k)\cdot({\bf v}\times({\bf S}^-_l+{\bf e}_l))\RA=(\Ae^-)^k_l.$$
By comparison of the above relations, $(\Ae^-)^k_l=\Ae^l_k$,
i.e., the $\alpha$-effect tensor $\A^-$ for the reverse flow $-\bf v$
is obtained from the tensor $\A$ for $\bf v$ by transposition. Consequently,
the maximum slow-time growth rates due to the action of the $\alpha$-effect
in the direct and reverse flows coincide.

\subsection{Magnetic eddy diffusivity}\label{medi}

An important non-generic case is that of the absence of the $\alpha$-effect,
i.e., when $\A=0$, whereby \rf{aleq} yields $\lambda_1=0$, but
$\LA{\bf b}_0\RA$ remains undetermined. This occurs, e.g., if the flow $\bf v$
is parity-invariant (a vector field $\bf f$ is parity-invariant as long as
${\bf f}(-{\bf x})=-{\bf f}({\bf x})$ and parity-antiinvariant when
${\bf f}(-{\bf x})={\bf f}({\bf x})$). For such flows, parity-invariant and
parity-antiinvariant fields constitute invariant subspaces of the magnetic
induction operator \rf{Leig}. Consequently, ${\bf S}_k({\bf x})$
are parity-antiinvariant implying $\A=0$.

From the second equation in the hierarchy we then find
$${\bf b}_1=\sum_{k=1}^3\sum_{m=1}^3{\partial\LA{\bf b}_0\RA_k\over\partial X_m}
\,{\bf G}_{mk},$$
where an appropriate normalisation of the magnetic mode $\bf b$ \rf{bex} is
assumed, and the small-scale zero-mean (not necessarily solenoidal!) fields
${\bf G}_{mk}({\bf x})$ are solutions to 9 {\it auxiliary problems of type~II}:
\BE\L{\bf G}_{mk}=-2\eta{\partial{\bf S}_k\over\partial x_m}
-{\bf e}_m\times({\bf v}\times({\bf S}_k+{\bf e}_k)).\EE{Geq}
When $\bf v$ is parity-invariant, ${\bf G}_{mk}({\bf x})$ are parity-invariant
as well. (Actually, for such $\bf v$ no odd powers of $\varepsilon$ enter
the series \rf{lex} for the eigenvalue $\lambda$, and in the expansion \rf{bex}
of the mode, ${\bf b}_n$ are parity-antiinvariant for all even indices
and parity-invariant for odd $n$, see section 3.5 of~\cite{VZ}.)

The mean magnetic mode $\LA{\bf b}_0\RA$ is a solution
to the eigenvalue problem for the operator of magnetic eddy diffusivity:
\BE\eta\nabla^2_{\bf X}\LA{\bf b}_0\RA+\nabla_{\bf X}\times\sum_{k=1}^3\sum_{m=1}^3\D_{mk}
{\partial\LA{\bf b}_0\RA_k\over\partial X_m}=\lambda_2\LA{\bf b}_0\RA,\EE{med}
which is the solvability condition for the third (order $\varepsilon^2$)
equation in the hierarchy. Here
\BE\D_{mk}=\LA{\bf v}\times{\bf G}_{mk}\RA\EE{Ddef}
is the so-called tensor of magnetic eddy diffusivity correction.
The expression \rf{Ddef} conveys the physically important idea that, like
the $\alpha$-effect, eddy diffusivity is a manifestation
of the interaction of fine structures of the flow and magnetic field, but due
to the parity it is order $\varepsilon$ weaker than the $\alpha$-effect
that is generically present in the absence of symmetries.

We solve \rf{med} for the mean mode \rf{mm}, following essentially the same
approach (see \cite{ABNZ}) as is used in section \ref{mae} for the problem
\rf{aleq}. Using the relations \rf{Bq}, we recast \rf{med}
into an eigenvalue problem for a $2\times2$ matrix:
\ba-\sum_{m,l,n}\De^l_{mn}q^{\rm p}_l(B_tq^{\rm t}_n+B_pq^{\rm p}_n)q_m
&=(\eta+\lambda_2)B_t,\\
\sum_{m,l,n}\De^l_{mn}q^{\rm t}_l(B_tq^{\rm t}_n+B_pq^{\rm p}_n)q_m&
=(\eta+\lambda_2)B_p.\end{align*}
Clearly, it transforms into \rf{Am} upon changing
$\eta+\lambda_2\to\lambda_1$ and $\sum_m\De^l_{mn}q_m\to-\I\Ae^l_n$. We
use this analogy to write down the solution:
\se{dei}\be\lambda_{2_\pm}({\bf q})&=-\eta-{1\over2}\sum_{j,l,n}
(D^l_n-D^n_l)q_j\pm\sqrt d,\label{d2}\\
d&=\sum_{j,l,n}\left(((\sD^l_n)^2-\sD^l_l\,\sD^n_n)q_j^2-2q_jq_n(\sD^l_n\,
\sD^l_j-\sD^l_l\,\sD^n_j)\right),\label{d3}\end{align}
where both sums are over even permutations of indices 1, 2 and 3 (i.e.,
$\epsilon_{jln}=1$) and it is denoted
\BE D^l_n=\sum_m\De^l_{mn}q_m,\qquad\sD^l_n=(D^l_n+D^n_l)/2.\EE{d1}\end{subequations}
As for the $\alpha$-effect operator, a compact expression
$$d=-{\bf q}\cdot(\det\mD)\,\mD^{-1}\bf q$$
follows from \rf{d3} if the matrix $\mD$ is invertible; however, this expression
does not help any more to determine the minimum
\BE\eta_{\rm eddy}\equiv\min_{|{\bf q}|=1}(-{\rm Re}\,\lambda_{2_\pm}({\bf q}))\EE{miE}
called the minimum magnetic eddy diffusivity, since by \rf{d1} $\mD$
depends on $\bf q$. If $d<0$, the magnetic mode
experiences oscillations in the slow time $T_2=\varepsilon^2t$; then, in view
of \rf{dei}, the frequency of oscillations is controlled
by the symmetric part, $^{\rm s}\D$, of the eddy diffusivity tensor, whose
entries are $(\De^l_{mn}+\De^n_{ml})/2$, and the slow-time growth or decay rate
of the mode is controlled by the antisymmetric part, $\D-{^{\rm s}\D}$.

Computing the eddy diffusivity correction tensor applying \rf{Ddef} requires
solving 12 auxiliary problems (3 of type I and 9 of type II), which is
numerically inefficient. A preferable alternative is to rely on the identity
\BE\De^l_{mk}=\LA{\bf Z}_l\cdot\left(2\eta{\partial{\bf S}_k\over\partial x_m}
+{\bf e}_m\times({\bf v}\times({\bf S}_k+{\bf e}_k))\right)\RA\EE{Dlmk}
expressing the entries of the tensor in terms of the solutions ${\bf S}_k$
to the 3 auxiliary problems of type~I and zero-mean solutions ${\bf Z}_l$
to 3 {\it auxiliary problems for the adjoint operator}:
\BE\L^*{\bf Z}_l={\bf v}\times{\bf e}_l\qquad\Leftrightarrow\qquad
\L^-(\nabla\times{\bf Z}_l+{\bf e}_l)=0,\EE{adj}
where
$$\L^*\,{\bf z}=\eta\nabla^2{\bf z}-{\bf v}\times(\nabla\times{\bf z})$$
is the operator adjoint to $\L$, and $\L^-$ is the magnetic induction operator
\rf{Lm} for the reverse flow $-\bf v$.
Unlike \rf{Ddef}, \rf{Dlmk} does not offer any evident physical interpretation.
By comparison of \rf{Seq} and~\rf{adj},
\BE\nabla\times{\bf Z}_l={\bf S}^-_l\quad\Rightarrow\quad{\bf Z}_l=
\eta^{-1}\nabla^{-2}({\bf v}\times({\bf S}^-_l+{\bf e}_l)).\EE{Zl}
Here $\nabla^{-2}$ denotes the inverse Laplacian in the fast variables
and ${\bf S}^-_l+{\bf e}_l$ is a neutral mode of $\L^-$, see \rf{Sml}.
When the small-scale dynamo does not operate (i.e., all
eigenvalues of $\L$ have non-positive real parts), the six fields
${\bf S}_k+{\bf e}_k$ and ${\bf S}^-_l+{\bf e}_l$ can be
computed as the small-scale dominant eigenmodes of the magnetic induction
operators $\L$ and $\L^-$, respectively; the same small-scale eigenvalue code
(e.g., \cite{Zh}) solves all the 6 eigenproblems (with the flow reversed,
$\bf v\to-v$, when computing ${\bf S}^-_l$).

\subsection{Dynamo powered by ``weak turbulence''}\label{hs}

The two-scale dynamos considered in this section thus far are characterised
by order 1 local magnetic Reynolds numbers R$_m^{\rm loc}=\ell\LA|{\bf v}|^2\RA^{1/2}\!\!/\eta$,
given that the small-scale flow $\bf v$, the size
of its periodicity box $\ell=2\pi$ and the molecular diffusivity $\eta$ are all
assumed to be non-dimensional and independent of the scale ratio $\varepsilon$
(in computations reported in sections \ref{al} and \ref{ed}, the flow has been
normalised and its r.m.s.~velocity is 1). The $\alpha$-effect tensor for dynamos
driven by turbulence was expressed in terms of the helicity spectrum
\cite{M70,MP}. We will now reproduce this result for small R$_m^{\rm loc}$ (this
modelling magnetic field generation by ``weak turbulence'') by the multiscale
asymptotic techniques. R$_m^{\rm loc}$ becomes small when either the molecular
eddy diffusivity is large, or when the amplitude of the flow is small. Let us
inquire into the two cases.

In the former case we consider the repeated limit
$\varepsilon\to0$ and $\eta\to\infty$. The first limit, in $\varepsilon$,
yields the asymptotic expansions \rf{ble}; we just need to calculate the second
one, in $\eta$. From \rf{Seq}, the neutral modes have the asymptotics
$${\bf S}_k=-\eta^{-1}\nabla^{-2}{\partial{\bf v}\over\partial x_k}
+{\rm O}(\eta^{-2}),$$
and hence the $\alpha$-effect tensor is composed of the columns \rf{Adef}
\BE\A_k=-\eta^{-1}\LA{\bf v}\times\nabla^{-2}{\partial{\bf v}\over\partial
x_k}\RA+{\rm O}(\eta^{-2}).\EE{as}

In the latter case a time-periodic velocity $\varepsilon^{1/N}{\bf v(x},t)$
is assumed; we thereby relax till the end of the section
the condition of steadiness of the flow, in order to obtain the expression
for the $\alpha$-effect tensor derived in \cite{MP} by the Test Field Method.
(Note that the multiscale
formalism discussed in the previous subsections has been generalised
to encompass dynamos, periodic in the fast time, see chapter 4 in \cite{VZ};
the present work is concerned with steady flows only, because they give rise
to significantly computationally less demanding auxiliary problems than
in the time-periodic setup.) Here $N>1$ is an integer. Following \cite{V86}
(see also \cite{GFP}), we solve the Floquet problem
$(-\partial/\partial t+\L){\bf b}=\lambda\bf b$ for large-scale
magnetic modes $\bf b$ of the same period $2\pi/\omega$ in the fast time $t$
as that of the flow. The modified expansions \rf{ble} now take the form
\se{blN}\be{\bf b}({\bf X,x},t)&=\sum_{n=0}^\infty{\bf b}_n({\bf X,x},t)\,\varepsilon^{n/N},\label{beN}\\
\lambda&=\sum_{n=0}^\infty\lambda_n\varepsilon^{n/N}.\label{leN}\end{align}\end{subequations}
Substituting them into the eigenvalue equation yields a hierarchy of equations
\BE-{\partial{\bf b}_n\over\partial t}
+\eta\nabla^2{\bf b}_n+\nabla\times({\bf v}\times{\bf b}_{n-1})
+2\eta(\nabla\cdot\nabla_{\bf X}){\bf b}_{n-N}+\nabla_{\bf X}\times
({\bf v}\times{\bf b}_{n-N-1})+\eta\nabla^2_{\bf X}{\bf b}_{n-2N}
=\sum_{m=0}^n\lambda_{n-m}{\bf b}_m\EE{hiN}
from which we can successively find all terms in the series \rf{blN}. For a steady
flow $\bf v$ and $N=2$, \rf{blN} determined by this procedure were proven
\cite{V86} to be asymptotic series for a multiscale solution to the eigenvalue
problem for the magnetic induction operator (provided the eigenvalue
$\lambda_4$ of the limit operator in the l.h.s.~of \rf{eq4} has multiplicity 1,
and the spatial periodicity of the eigenfunction is compatible with that
of the flow); the proof can be recast for time-periodic flows.

Averaging over the space-time periodicity cell (we will denote this average
by double angle brackets, $\langle\!\langle\ \rangle\!\rangle$) order
$\varepsilon^{n/N}$ equations for $n\le N+1$, we find $\lambda_n=0$ for all
such $n$, provided $\langle\!\langle{\bf b}_0\rangle\!\rangle$ does not vanish
identically. The order $\varepsilon^0$ equation yields
${\bf b}_0={\bf b}_0({\bf X})$. The order $\varepsilon^{1/N}$ equation is then
equivalent to
\BE{\bf b}_1=(\partial/\partial t-\eta\nabla^2)^{-1}({\bf b}_0\cdot\nabla)\bf v,\EE{b1}
where the inverse operator is calculated in the space of vector fields,
$2\pi$-periodic in space and $2\pi/\omega$-periodic in the fast time, whose
spatio-temporal mean vanishes. Consequently, the order $\varepsilon^{(N+2)/N}$
equation becomes upon averaging a closed eigenvalue equation in ${\bf b}_0$:
\se{lim}\be\eta\nabla^2_{\bf X}{\bf b}_0+\nabla_{\bf X}\times(\A{\bf b}_0)
=&\,\lambda_4{\bf b}_0\phantom{\lambda_{N+2}}\mbox{for}~N=2;\label{eq4}\\
\nabla_{\bf X}\times(\A{\bf b}_0)
=&\,\lambda_{N+2}{\bf b}_0\phantom{\lambda_4}\mbox{for}~N>2,\label{eqN}\end{align}\end{subequations}
where the $\alpha$-effect tensor consists of columns
\BE\A_k=\left\langle\hspace*{-1.414ex}\left\langle{\bf v}\times\left({\partial\over\partial t}
-\eta\nabla^2\right)^{-1}{\partial{\bf v}\over\partial x_k}\right\rangle
\hspace*{-1.414ex}\right\rangle.\EE{Anew}

In the two cases under consideration, the limit R$_m^{\rm loc}\to0$
is approached along different paths in the parameter space; thus, it is
no wonder that the equation \rf{eq4} for determining the leading terms
in the series \rf{blN} differs from \rf{aleq} and \rf{eqN}.
Nevertheless, the $\alpha$-effect tensors are the same!
(For steady flow, \rf{Anew} clearly reduces to \rf{as}; in fact, calculations
for a time-periodic flow in the first case yield an expression identical
to \rf{Anew}, up to an O($\eta^{-2}$) discrepancy.)

Following \cite{M70,MP}, we expand the velocity in a Fourier series,
$${\bf v(x},t)=\sum_{{\bf p},p'}\hat{\bf v}_{{\bf p},p'}\,
\e^{\I(p'\omega t+\bf p\cdot x)}$$
and transform \rf{Anew}:
$$\A_k=\sum_{{\bf p},p'}{\I p_k\over\I p'\omega+\eta|{\bf p}|^2}
\,(\overline{\hat{\bf v}}_{{\bf p},p'}\times\hat{\bf v}_{{\bf p},p'}).$$
Here summation is over all integer-component four-dimensional vectors
$({\bf p},p')\ne0$ and the bar denotes complex conjugation. The contribution
of two terms with opposite indices, ${\bf p},p'$ and $-{\bf p},-p'$, is real.
Since both factors in the vector product in the r.h.s.~are normal to $\bf p$,
the product is parallel to $\bf p$, and hence finally
$$\A_k=-\sum_{{\bf p},p'}{\eta p_k{\bf p}\over p'^2\omega^2+\eta^2|{\bf p}|^4}
H_{{\bf p},p'}.$$
Here
\BE H_{{\bf p},p'}=\overline{\hat{\bf v}}_{{\bf p},p'}\cdot
(\I{\bf p}\times\hat{\bf v}_{{\bf p},p'}),\EE{hm}
called the helicity spectrum of the flow $\bf v$, is the set of Fourier
coefficients of the convolution integral ${\cal C}({\bf r},\tau)=\langle\!
\langle{\bf v(x},t)\cdot(\nabla\times{\bf v(x+r},t+\tau))\rangle\!\rangle$.
Evidently, ${\cal C}(0,0)$ is the mean kinetic helicity. This only
link between kinetic helicity and the helicity spectrum is loose: in general,
pointwise vanishing of the kinetic helicity density \rf{non} neither requires,
nor implies vanishing of the helicity spectrum. We observe that for
R$_m^{\rm loc}\to0$ no $\alpha$-effect is present if $H_{\bf p}=0$ for all $p$.

Suppose now the flow is parity-invariant and thus $\A=0$. As above for
the $\alpha$-effect tensor, we first consider the limit of small local magnetic
Reynolds numbers realised as a repeated limit $\varepsilon\to0$ and
$\eta\to\infty$. The evolutionary versions of \rf{Seq} and \rf{Geq} imply
$${\bf S}_k={\partial{\bf u}\over\partial x_k}+{\rm O}(\eta^{-2}),\qquad
{\bf G}_{mk}=2\eta\left({\partial\over\partial t}-\eta\nabla^2\right)^{-1}
{\partial^2{\bf u}\over\partial x_k\partial x_m}
+{\bf e}_m\times({\bf u}\times{\bf e}_k)+{\rm O}(\eta^{-2}),$$
where
\BE{\bf u}=(\partial/\partial t-\eta\nabla^2)^{-1}{\bf v}.\EE{u}
By \rf{Ddef}, where spatio-temporal averaging is assumed as required
for time-periodic fields,
$$\D^l_{mk}=\epsilon_{nlk}\sum_{({\bf p},p')\ne0}
{\overline{\hat v}\vphantom{\bf v}_{{\bf p},p'}^n\hat v_{{\bf p},p'}^m
\over\I p'\omega+\eta|{\bf p}|^2}+{\rm O}(\eta^{-2}),$$
where $n=6-l-k$ for $l\ne k$. This eddy diffusivity tensor corresponds to
the ``symmetric part of $\beta_{iml}$'' in \cite{MP}. Since $\mD$ vanishes
in the leading order, $\lambda_{2_\pm}({\bf q})$ \rf{d2} coincide up to
an O$(\eta^{-2})$ discrepancy:
$$\lambda_{2_\pm}({\bf q})=-\eta-\eta\sum_{({\bf p},p')\ne0}
{|{\bf p}|^2|{\bf q}\cdot\hat{\bf v}_{{\bf p},p'}|^2
\over p'^2\omega^2+\eta^2|{\bf p}|^4}+{\rm O}(\eta^{-2}).$$
Hence, in this case eddy correction of magnetic diffusivity is predominantly
positively defined and only enhances molecular diffusivity. (This is
reminiscent of passive scalar transport \cite{Bi}.) The ``antisymmetric
part of $\beta_{iml}$'' of \cite{MP} linked with the helicity spectrum does not
show up in the leading terms of the asymptotic expansion.

In the second case, the flow $\varepsilon^{1/N}{\bf v(x})$ again forces
expanding large-scale magnetic modes and associated eigenvalues
in the power series \rf{blN} in $\varepsilon^{1/N}$. All terms can be determined
from the hierarchy \rf{hiN}. As in the presence of the $\alpha$-effect,
${\bf b}_0={\bf b}_0({\bf X})$ and ${\bf b}_1$ satisfies \rf{b1}.
For a parity-invariant flow, ${\bf b}_n$ are parity-antiinvariant for $n\le N$,
but gain parity-invariant parts for larger $n$. Averaging yields $\lambda_n=0$
for all $n<2N$. The parity-invariant part of ${\bf b}_{N+1}$,
$$-2\eta(\nabla\cdot\nabla_{\bf X})({\bf b}_0\cdot\nabla)({\partial/\partial t}
-\eta\nabla^2)^{-1}{\bf u}-({\bf u}\cdot\nabla_{\bf X}){\bf b}_0,$$
where $\bf u$ is defined by \rf{u}, does give rise to a mean e.m.f., but
only at the order $\varepsilon^{(2N+2)/N}$ equation. However, a closed equation
for determining ${\bf b}_0$ and $\lambda_{2N}$ emerges at an earlier stage
as a solvability condition for the order $\varepsilon^2$ equation; it is just
an eigenvalue problem for the Laplace operator in slow variables, i.e.,
in the second case no correction of magnetic diffusivity due to small-scale
fields arises in the leading order.

We conclude that the multiscale asymptotic analysis does not confirm that
magnetic eddy diffusivity is linked with the helicity spectrum of the flow.

\section{Construction of zero-helicity flows}\label{const}

In this section we present some approaches for construction
of three-dimensional pointwise non-helical flows. For convenience of reference,
we will categorise such flows by the techniques applied for constructing them;
flows that are obtained by a specific technique will be said to constitute
{\it a~family}. We consider six different families. The classification
is imprecise in that the families may have non-trivial intersections.

\subsection{Poloidal flows: family P}\label{fP}

We discuss here the semianalytical construction of poloidal non-helical flows,
which will be called {\it family P}.
The poloidal flow for the potential $P({\bf x})$ is
\BE{\bf v}=\left({\partial^2 P\over\partial x_1\partial x_3},\
{\partial^2 P\over\partial x_2\partial x_3},\ -\nabla^2_{x_1x_2}P\right),\EE{pol}
where
$$\nabla^2_{x_1x_2}={\partial^2\over\partial x_1^2}
+{\partial^2\over\partial x_2^2}$$
is the Laplacian in the horizontal coordinates $x_1,\,x_2$. The kinetic
helicity density of the flow \rf{pol} vanishes as long as
\BE{\partial\nabla^2P\over\partial x_2}\ {\partial^2P\over\partial
x_1\partial x_3}-{\partial\nabla^2P\over\partial
x_1}\ {\partial^2 P\over\partial x_2\partial x_3}=0.\EE{polz}
We regard \rf{polz} as a first-order equation in $\nabla^2P$
and tackle it by the method of characteristics.
Characteristics $(x_1(\tau),x_2(\tau),x_3(\tau))$ satisfy the ODEs
$$\dot{x}_1=-\partial^2 P/\partial x_2\partial x_3,\qquad
\dot{x}_2=\partial^2 P/\partial x_1\partial x_3,\qquad\dot{x}_3=0,$$
which is equivalent to $\partial P/\partial x_3=\rm constant$ and
$x_3=\rm constant$, i.e., the characteristics are categorised by the common,
along a characteristic, marker values of $\partial P/\partial x_3$ and
vertical coordinate~$x_3$. Then \rf{polz} states that along
a characteristic $\nabla^2P$ does not vary. Therefore,
$\nabla^2P$ depends only on the marker values:
\BE\nabla^2P=F\left(\partial P/\partial x_3,x_3\right),\EE{pole}
where $F$ is an arbitrary function of two scalar arguments.

In principle, we can select a smooth function $F$ and attempt to solve \rf{pole}
numerically for the potential $P$ in $\T^3$. A particular case where
the dependence of $P$ on $x_3$ is restricted to a multiplicative one,
\BE P({\bf x})=g(x_1,x_2)\,\widetilde{g}(x_3),\EE{pot}
is significantly simpler than the general problem \rf{pole}.
Now $g$ must satisfy an equation
\BE\nabla^2_{x_1x_2}\,g=f(g)\EE{red}
in a planar cell of periodicity $\T^2$, so that \rf{pot} satisfies \rf{pole} for
$F(D,x_3)=\widetilde{g}f(D/\widetilde{g}\,')+D\,\widetilde{g}\,''/\widetilde{g}\,'$,
where prime mark $'$ denotes a derivative in $x_3$. In the problem \rf{red},
we may set
\BE f(g)=\sum_{j=1}^J\nu_j(f_j(g)-\LA f_j(g(x_1,x_2))\RA).\EE{fi}
Here $\widetilde{g}(x_3)$ and $f_j(g)$ are arbitrary smooth functions.
The mean of the r.h.s.~is removed, because the l.h.s.~of \rf{red} is zero-mean;
as a result, when projected onto a finite-dimensional Fourier subspace,
\rf{red} yields one equation less than the number of harmonics involved
in the approximate solution. This can be used to enforce the condition that
the unknown function $g(x_1,x_2)$ is zero-mean, which helps to bypass
the emergence of constant-value solutions. ``Eigenvalues'' $\nu_j$
in \rf{fi} can be calculated as constants minimising the discrepancy.

We solve the problem \rf{red}--\rf{fi}
numerically by a quasi-newtonian procedure (see \cite{NR}); at each iteration,
the respective linear problem is solved by an optimised version of the BiCGstab
method \cite{Fok,SF,SV95,SV96}. (We apply BiCGstab($\ell$) for $2\le\ell\le7$
to generate a sequence of ``raw'' approximations to the solution. The best
approximation known so far is stored, and each time BiCGstab has computed new
$K$ raw approximations, their optimal linear combination is used to improve
the best approximation.)

For $J=1$, substituting \rf{fi} transforms \rf{red} into a kind of nonlinear
eigenvalue problem (see, e.g., \cite{Poh}). For $f_1(g)=g$ it reduces to
the standard eigenvalue problem for the Laplace operator; this yields
analytical examples of non-helical poloidal flows.
The Christopherson flow \rf{Chris} falls into this category (albeit
the periodicity box of \rf{Chris} is a parallelepiped distinct from a cube):
its poloidal potential
$$P({\bf x})={L^2\over4\pi^2}\,g(x_1,x_2)\sin\pi x_3$$
satisfies \rf{pot}--\rf{red} for $f(g)=-4\pi^2(3L^2)^{-1}g$.

\subsection{Application of the Monge decomposition; family L}\label{fL}

Note that
\BE\nabla A\times\nabla B=\nabla\times(A\nabla B)=
{1\over2}\,\nabla\times(A\nabla B-B\nabla A)\EE{curl}
is a solenoidal field. Conversely, any solenoidal vector field can be
locally expressed as \rf{curl} (called the Monge decomposition)
in terms of two scalar functions $A({\bf x})$ and $B({\bf x})$ \cite{Ph}
known as Clebsch variables \cite{Lamb} or Monge potentials \cite{Tru}.

A vector field
\se{vf}\be{\bf v(x)}&=A\nabla B-\nabla p\label{v1}\\
&=(A\nabla B-B\nabla A)/2-\nabla p_\star\label{v2}\end{align}\end{subequations}
is solenoidal as long as
\be\nabla^2 p&=\nabla A\cdot\nabla B+A\nabla^2 B,\label{p1}\\
\nabla^2p_\star&=(A\nabla^2 B-B\nabla^2 A)/2\label{p2}\end{align}
(whereby $p-p_\star=AB/2$). We seek non-helical solenoidal vector fields
\rf{vf}, whose Monge potentials $A({\bf x})$ and $B({\bf x})$
are defined in the entire torus $\T^3$. By virtue of \rf{curl}, a field
\rf{vf} is pointwise non-helical provided
\BE\nabla p\cdot(\nabla A\times\nabla B)=\nabla p_\star\cdot(\nabla A\times\nabla B)=0\EE{hell}
globally (see the discussion of the so-called ``complex-lamellar flows'' in
\cite{Tru}).

Actually, the condition \rf{hell} is equivalent to demanding that, at least
locally, $p({\bf x})=\widetilde{p}(A({\bf x}),B({\bf x}))$ as well as
$p_\star({\bf x})=\widetilde{p}_\star(A({\bf x}),B({\bf x}))$.
By the chain rule, for such a $p$ \rf{hell} holds true.
To show the converse, note that a field \rf{vf} vanishes identically
unless $A$ and $B$ are functionally independent (since otherwise $A\nabla B$
is a gradient). Thus, we can express $p$ in some local coordinates
$(A({\bf x}),B({\bf x}),C({\bf x}))$ so that
$$\nabla\widetilde{p}\cdot(\nabla A\times\nabla B)=
\nabla C\cdot(\nabla A\times\nabla B)\,\partial\widetilde{p}/\partial C,$$
and therefore $\partial\widetilde{p}/\partial C=0$, which implies the statement.

By virtue of \rf{p2}, a flow \rf{v2} is solenoidal and non-helical for $p_\star=0$
when $A$ and $B$ are eigenfunctions of the Laplace operator, $\nabla^2$,
(or, moreover, of the operator $\rho\nabla^2$, where $\rho({\bf x})$ is
an arbitrary function)
that are associated with the same eigenvalue. (Since the multiplicity of most
eigenvalues of the Laplacian in $\T^3$ exceeds 1, such independent functions
$A$ and $B$ do exist.) Such flows constitute {\it family L}. Further
examples of non-helical flows that will be presented in this section also rely
on the Monge decomposition.

\subsection{Cosine flows: family C}\label{fC}

We call {\it family C} or {\it the cosine flows} the solenoidal non-helical
flows defined as
\be v_1&=n(b_1\sin({\bf a\cdot x})+a_1\sin({\bf b\cdot x}))\cos nx_3,\nonumber\\
v_2&=n(b_2\sin({\bf a\cdot x})+a_2\sin({\bf b\cdot x}))\cos nx_3,\label{cos}\\
v_3&=-({\bf a\cdot b})(\cos({\bf a\cdot x})+\cos({\bf b\cdot x}))\sin nx_3,
\nonumber\end{align}
where ${\bf a}=(a_1,a_2,0)$ and ${\bf b}=(b_1,b_2,0)$ are constant horizontal
vectors.

The field \rf{cos} is obtained from \rf{v2} for
\ba A&=|\cos({\bf(b+a)\cdot x}/2)|^\alpha\ |\cos({\bf(b-a)\cdot x}/2)|^\beta\ |\cos nx_3|^\chi,\\
B&={4n\over A(\alpha-\beta)}\cos({\bf(b+a)\cdot x}/2)
\cos({\bf(b-a)\cdot x}/2)\cos nx_3\end{align*}
(hence the name of the flows), where $\alpha,\,\beta$ are arbitrary constants and
\be\chi=(({\bf a\cdot b})(\beta-\alpha)+n^2(\beta+\alpha))/(2n^2).
\label{chi}\end{align}
In particular, for
$$\alpha={1\over2}\left(1+{n^2(1-2\chi)\over\bf a\cdot b}\right)\quad\mbox{and}
\quad\beta={1\over2}\left(1-{n^2(1-2\chi)\over\bf a\cdot b}\right),$$
which implies \rf{chi}, \rf{v2} is solenoidal for $p_\star=0$ and hence
non-helical. (As a side remark, note that this example demonstrates
non-uniqueness of the fields $A({\bf x}),\,B({\bf x}),\,p_\star({\bf x})$
realising a flow \rf{v2}.)

The cosine flows \rf{cos} have non-zero toroidal, $T$, and poloidal, $P$, potentials
\ba T({\bf x})&=n(a_1b_2-a_2b_1)(|{\bf a}|^{-2}\cos({\bf a\cdot x})
-|{\bf b}|^{-2}\cos({\bf b\cdot x}))\cos nx_3,\\
P({\bf x})&=-({\bf a\cdot b})(|{\bf a}|^{-2}\cos({\bf a\cdot x})
+|{\bf b}|^{-2}\cos({\bf b\cdot x}))\sin nx_3.\end{align*}
Consequently, the Christopherson flow \rf{Chris} does not belong to this family.

\subsection{An eigenfunction approach; family I}\label{fI}

Constructing a non-helical flow \rf{v1} for a prescribed smooth Monge potential
$B({\bf x})$ requires finding such
an $A({\bf x})$ that the space-periodic $p$, uniquely determined from \rf{p1},
satisfies \rf{hell}. In the Lebesgue space of scalar functions in $\T^3$,
which have a zero spatial mean, we define a pseudodifferential operator
\BE\M A=\nabla^{-2}(\nabla A\cdot\nabla B+A\nabla^2 B),\EE{oM}
where $\nabla^{-2}$ denotes the inverse Laplace operator (it is applied
in the l.h.s.~of this relation to a zero-mean field $\nabla\cdot(A\nabla B)$;
the result has a zero mean by the definition of the inverse Laplacian).

By standard arguments, $\M$ is a compact operator, whose eigenfunctions
not belonging to the kernel
are smooth. Any eigenfunction of $\M$ associated with a real eigenvalue
is a solution to our problem: by comparison of the eigenvalue equation
\BE\M A=\mu A\EE{eig}
with \rf{p1}, $p=\mu A$, which clearly satisfies \rf{hell}.
We must show that the eigenfunction of $\M$ is functionally independent
of $B({\bf x})$. Suppose the converse is true, i.e., $A({\bf x})=\widetilde{A}(B({\bf x}))$.
Substituting such an $A$ into \rf{eig} yields
$$\int_{B_0}^B\widetilde{A}(b)\d b=\mu(\widetilde{A}(B)-\widetilde{A}(B_0)),$$
where $B_0$ is a constant from the image of $B({\bf x})$. Differentiating
this equation in $B$ and solving the resultant ODE, for $\mu\ne0$ we find
$\widetilde{A}(B)=\widetilde{A}_0\e^{B/\mu}$, which has a zero spatial mean
only for $\widetilde{A}_0=0$; if $\mu=0$, then again $\widetilde{A}=0$.
This completes the demonstration.

The adjoint operator for $\M$ is
$$\M^*A^*=-\nabla(\nabla^{-2}A^*)\cdot\nabla B-\LA A^*B\,\RA.$$
Suppose the ODE $\dot{\bf x}=\nabla B$ has a global space-periodic first
integral $I({\bf x})$. Then, clearly, field $A^*=\nabla^2I$ belongs
to the kernel of $\M^*$ and hence the kernel of $\M$ is also non-empty.
Therefore for such a Monge potential $B$ there exists a solenoidal non-helical
flow \rf{v1} for $p=0$. Flows \rf{v1} whose existence is established
by this argument are designated {\it family I}.

Unfortunately, as it is shown in the next subsection, all eigenfunctions
of $\M$ that do not belong to its kernel are complex-valued and thus
are unsuitable for our purposes.

\subsection{Variable-separated flows: families V$_1$ and V$_2$}\label{fV}

Consider an equation
\BE\nabla\cdot(A\nabla B)=\mu\nabla^2(AB),\EE{mo}
whose solutions are Monge potentials of solenoidal non-helical flows \rf{v1} for
\BE p=\mu AB.\EE{p}
Unlike \rf{eig}, \rf{mo} is homogeneous in both $A$ and $B$. Let us derive its
variable-separated solutions. For $A({\bf x})=\prod_{i=1}^3A_i(x_i)$
and $B({\bf x})=\prod_{i=1}^3B_i(x_i)$, \rf{mo} transforms to
\BE{\d\ \over\d x_i}\left(\mu\,{\dot{A}_i\over A_i}+(\mu-1)\,{\dot{B}_i\over B_i}\right)
+\left(\mu\,{\dot{A}_i\over A_i}+(\mu-1)\,{\dot{B}_i\over B_i}\right)
\left({\dot{A}_i\over A_i}+{\dot{B}_i\over B_i}\right)=C_i,\EE{ode}
where the dot denotes differentiation in $x_i$ and $C_i$ are constants such that
\BE\sum_{i=1}^3C_i=0.\EE{cn1}
Regarded as a first-order linear ODE, \rf{ode} has a solution
\BE\mu\,{\dot{A}_i\over A_i}+(\mu-1)\,{\dot{B}_i\over B_i}={1\over A_iB_i}
\left(C_{2i}+C_i\int_0^{x_i}A_i(\xi)B_i(\xi)\,\d\xi\right),\EE{fst}
where $C_{2i}$ is a constant.
It is natural to solve \rf{fst} for a prescribed product
$$F_i(x_i)=A_iB_i,$$
since then the constants $C_i$ and $C_{2i}$ are the only unspecified data
in the r.h.s.~of \rf{fst}. Integration of \rf{fst} yields
\BE B_i=C_{3i}\,F_i^\mu(x_i)\,\exp\left(-\int_0^{x_i}\left(C_{2i}+C_i\int_0^\zeta
F_i(\xi)\,\d\xi\right){\d\zeta\over F_i(\zeta)}\right),\qquad A_i=F_i/B_i,\EE{AB}
where $C_{3i}$ is a constant that is irrelevant and will be set to unity.

If the constants $C_i$ and $C_{2i}$ are scaled by $\mu\ne0$, the flow
\rf{v1}, \rf{AB} turns out to be proportional to this parameter and not
to involve it otherwise. Thus, the value of $\mu\ne0$ is irrelevant for our
constructions. By virtue of \rf{p}, the flow \rf{v1} for $\mu=1$ takes the form
\BE{\bf v(x)}=-B\nabla A,\EE{v3}
whereby the case $\mu=0$ is also reduced to the generic case
$\mu=1$ essentially by swapping the Monge potentials $A$ and $B$.
Substituting \rf{AB} into \rf{v3}, we find the general form of solenoidal
non-helical flows constituting {\it family V$_1$} that are obtained by separation
of variables in the Monge potentials satisfying \rf{mo}:
\BE{\bf v(x)}=\left(C_1U_1\dot U_2\dot U_3,\ C_2\dot U_1U_2\dot U_3,\ C_3\dot U_1\dot U_2U_3\right).\EE{V1}
Here $C_i$ are arbitrary constants satisfying \rf{cn1} and
$U_i=-C_{2i}/C_i-\int_0^{x_i}F_i(\xi)\,\d\xi$ are arbitrary smooth
$2\pi$-periodic functions of $x_i$. When some $C_i=0$,
\rf{V1} is a planar flow that, by the Zeldovich \cite{Zel} antidynamo theorem,
cannot generate magnetic field and hence it is not of our interest.

Constructing more general solutions to \rf{mo} is difficult. Let now
$B({\bf x})$ be specified. For $B>0$, \rf{mo} can be regarded as
an eigenvalue problem for the compact operator
\BE A\mapsto(1/B)\nabla^{-2}(\nabla A\cdot\nabla B+A\nabla^2 B).\EE{oS}
However, for such a $B$ either an eigenvalue $\mu\ne0$ is complex, resulting
in a physically irrelevant complex field \rf{v1}, \rf{p}, or the associated
eigenfunction $A$ yields a zero flow. To show this, we substitute
$A=\widehat AB^{\chi-1}$, $B=\e^{\widehat B}$, where $\chi=1/(2\mu)$,
thereby reducing \rf{mo} to
$$\nabla^2\widehat A=(\chi\nabla^2\widehat B+\chi^2|\nabla\widehat B|^2)
\widehat A.$$
Multiplying this equation by $\widehat A$ and integrating over the periodicity
cell we obtain
$$\int_{\T^3}(|\nabla\widehat A|^2+\chi^2|\nabla\widehat B|^2|\widehat A|^2)\,\d{\bf x}
=2\chi\int_{\T^3}\widehat A\,\nabla\widehat A\cdot\nabla\widehat B\,\d{\bf x}.$$
For real $\widehat A$, $\widehat B$ and $\chi$, the r.h.s.~of this relation
does not exceed in absolute value the l.h.s., the equality takes place only
for $\nabla\widehat A=\chi\widehat A\nabla\widehat B$ and thus
$\widehat A=C\e^{\chi\widehat B}$, where $C$ is a constant. Consequently,
$A=CB^{2\chi-1}$, and the respective flow \rf{v1}, \rf{p} is zero.

We should therefore focus on solving \rf{mo} for $B$ that change sign
in the periodicity cell. For such a $B$, the operator \rf{oS} is singular,
which renders the problem \rf{mo} difficult for numerical treatment.

The eigenvalue problem \rf{eig} for the operator $\M$ \rf{oM}, which lacks
the problematic factor $1/B$, has the same terminal drawback as \rf{mo}:
Letting $B=\ln\widehat B$ and $\widehat A=A/\widehat B$, where $\widehat B>0$,
transforms \rf{eig} into a problem whose structure is identical to \rf{mo},
making the above arguments applicable to the problem \rf{eig}, \rf{oM}. Thus,
flows with the desirable properties can involve as a Monge potential $A$
only those eigenfunctions of $\M$ that belong to its kernel.

If we drop the condition that the flow complies with \rf{v1}, but still
demand that each its component is variable-separated, we encounter
{\it family V$_2$} of variable-separated solenoidal non-helical flows:
\BE{\bf v(x)}=(C_1U_2U_3,\ C_2U_1U_3,\ C_3U_1U_2).\EE{V2}
Here $C_i$ are arbitrary constants and $U_i$ arbitrary $2\pi$-periodic
smooth functions of $x_i$ (some of which should be zero-mean
to ensure that the mean velocity is zero).

When any $C_i$ vanishes, the flow \rf{V2} is planar; since by the Zeldovich
\cite{Zel} theorem such flows can not be dynamos, we do not consider them.
Examination of the product of the three scalar relations
$v_i(-{\bf x})=-v_i({\bf x})$ defining parity invariance of \rf{V2} reveals
that no family V$_2$ flow with all $C_i\ne0$ is parity-invariant.

\subsection{Helicity spectrum of the non-helical flows}\label{hsf}

We show here that the helicity spectrum of flows comprising families P, C,
V$_1$ and V$_2$ is identically zero; for family L flows this does not,
in general, hold true.

Vorticity of a poloidal flow \rf{pol} is
$$\nabla\times{\bf v}=\left(-{\partial\nabla^2 P\over\partial x_2},
{\partial\nabla^2 P\over\partial x_1},0\right).$$
Thus, integration by parts establishes ${\cal C}({\bf r})=0$. This
proves the claim for any poloidal flow, including family P flows.

It is straightforward to transform \rf{hm} using the reality of the flow $\bf v$:
$$H_{\bf m}=2{\bf m}\cdot({\rm Re}\,{\bf v}_{\bf m}
\times{\rm Im}\,{\bf v}_{\bf m})$$
(we have dropped the second index $m'$ not needed for steady flows).
Hence, all $H_{\bf m}=0$ when $\bf v$ is parity-invariant
(Re$\,{\bf v}_{\bf m}=0$) or parity-antiinvariant (Im$\,{\bf v}_{\bf m}=0$).
In particular, family C flows have zero helicity spectrum, as well as all
flows, for which the magnetic eddy diffusivity tensor is computed in section
\ref{ed}.

To consider flows from families V$_1$ and V$_2$, we expand the constitutive
functions in the Fourier series,
$$U_n(x_n)=\sum_{m_n}\hat{U}_{n,m_n}\e^{\I m_nx_n},$$
whereby a family V$_1$ flow has Fourier coefficients
$$\hat{\bf v}_{\bf m}=-\hat{U}_{1,m_1}\hat{U}_{2,m_2}\hat{U}_{3,m_3}
(C_1m_2m_3,C_2m_1m_3,C_3m_1m_2).$$
Thus, $H_{\bf m}=0$ since $\hat{\bf v}_{\bf m}$ and
$\hat{\bf v}_{-\bf m}$ are parallel.

A family V$_2$ flow has Fourier coefficients
$$\hat{\bf v}_{\bf m}=(C_1\delta_{m_1}^0\hat{U}_{2,m_2}\hat{U}_{3,m_3},
C_2\delta_{m_2}^0\hat{U}_{1,m_1}\hat{U}_{3,m_3},
C_3\delta_{m_3}^0\hat{U}_{1,m_1}\hat{U}_{2,m_2}),$$
where $\delta^j_i$ denotes the Kronecker symbol, and by straightforward
calculation of the determinant \rf{hm},
\BE H_{\bf m}=\I\sum_{1\le l,j,k\le 3}\epsilon_{ljk}
C_l\delta_{m_l}^0\overline{\hat{U}}_{j,m_j}\overline{\hat{U}}_{k,m_k}
m_jC_k\delta_{m_k}^0\hat{U}_{l,m_l}\hat{U}_{j,m_j}.\EE{FV2}
Flows that we consider are zero-mean, implying that
$\delta_{m_n}^0\!\hat{U}_{n,m_n}=0$ for at least two distinct indices~$n$.
Hence at least one such product vanishes in each of the six terms in \rf{FV2}.
Therefore for any zero-mean family V$_2$ flow the helicity spectrum also
vanishes.

Let us finally compute the helicity spectrum of a family L flow \rf{v2} for
$$A=\sum_{\bf n}\hat{A}_{\bf n}\e^{\I\bf n\cdot x},\qquad
B=\sum_{\bf n}\hat{B}_{\bf n}\e^{\I\bf n\cdot x}.$$
This implies
$$\hat{\bf v}_{\bf m}={\I\over2}\sum_{\bf n}\hat{A}_{\bf m-n}\hat{B}_{\bf n}
(2{\bf n-m}),$$
whereby
$$H_{\bf m}=\I\sum_{\bf n,j}\overline{\hat{A}}_{\bf m-n}\overline{\hat{B}}_{\bf n}
\hat{A}_{\bf m-j}\hat{B}_{\bf j}({\bf n}\cdot({\bf m}\times{\bf j})).$$
Generically, these quantities do not vanish.

\section{Magnetic $\alpha$-effect in non-helical flows: numerical results}\label{al}

We explore here the $\alpha$-effect featured by some non-parity-invariant
sample flows
belonging to families P, V$_1$, V$_2$ and L (see sections \ref{fP}, \ref{fV} and \ref{fL}).
The cosine flows are not considered in this section, since the $\alpha$-effect
tensor vanishes for a parity-invariant flow, and they have this symmetry.

We focus on the maximum real part $\gamma_\alpha$ \rf{mgr} of eigenvalues
of the $\alpha$-effect operators, and in section~\ref{ed} on the minimum
magnetic eddy diffusivity $\eta_{\rm eddy}$ \rf{miE}. For each pair (a flow /
molecular magnetic diffusivity value) employed, we have also computed the fast-time
growth rates $\gamma_{\rm sm}$ of dominant small-scale magnetic modes.
(A small-scale magnetic mode is an eigenfunction of the magnetic
induction operator $\L$, that has the same periodicity as the flow, i.e., in our
work, modes whose periodicity cell is $\T^3$. A mode is dominant, when it has
the maximum growth rate, i.e., the maximum real part of the associated
eigenvalue, over all modes residing in the same periodicity box, for the same
flow and molecular diffusivity.) We are especially interested in eigenvalues
of the $\alpha$-effect operator with strictly positive real parts and
in negative eddy diffusivity (both imply generation of large-scale magnetic
field) for pairs (a flow / molecular diffusivity) such that no generation
of small-scale modes occurs: While the fast-time growth rates of magnetic fields
generated by the $\alpha$-effect are order $\varepsilon$ and by the action
of negative eddy diffusivity order $\varepsilon^2$, growth rates
of small-scale modes are order $\varepsilon^0$. Thus, the $\alpha$-effect and
negative eddy diffusivity are significantly weaker mechanisms for generation
of large-scale magnetic field than small-scale dynamo,
and they gain importance only in the absence of small-scale generation.

Numerical work reported in this and the next sections has much in common.
Computation of dominant small-scale magnetic modes and their fast-time growth
rates,
as well as of solutions to auxiliary problems of type I and for the adjoint
operator (when needed for computing the eddy diffusivity tensor) has been
performed using the code \cite{Zh}.
Pseudo-spectral methods have been applied, typically, with the resolution
of $128^3$ Fourier harmonics. For validation of results, computations have been
repeated with the double resolution of $256^3$ harmonics for the smallest
magnetic molecular diffusivity used to analyse dependencies
of various quantities on the molecular diffusivity, or for several ``typical''
flows if the molecular diffusivity was not varied in a series of runs.
In these test runs, the results of $128^3$ harmonics computations have always
been confirmed to at least $10^{-7}$; thus even the smallest, order $10^{-5}$
values reported here have at least 2 significant digits.

Upon construction each flow, for which results are reported
in the present or next section, has been normalised before proceeding
with the dynamo computations, so that its r.m.s.~velocity is 1.

\subsection{Family P}\label{alP}

\begin{figure}[p]
\centerline{\raisebox{2.3in}{(a)} \ \raisebox{6mm}{\includegraphics[width=3in]{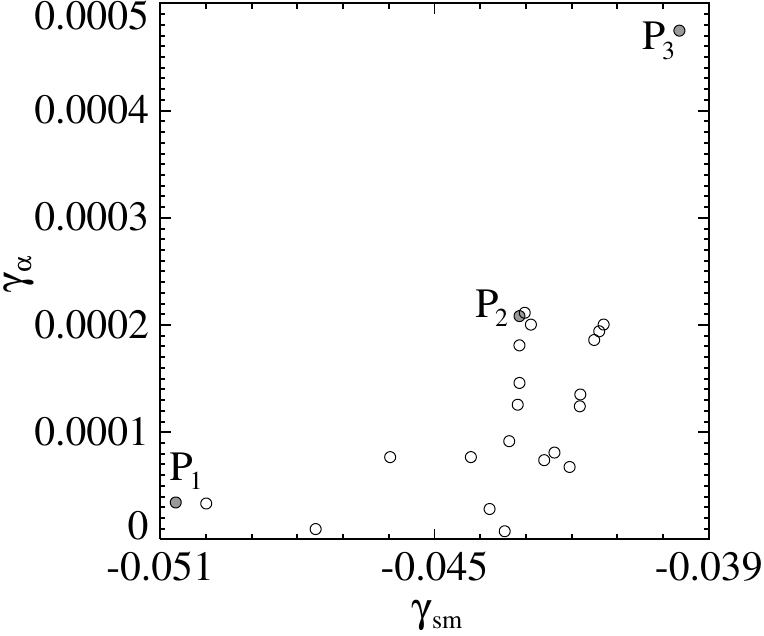}}
\hfill\raisebox{2.3in}{(b)} \ \includegraphics[width=3in]{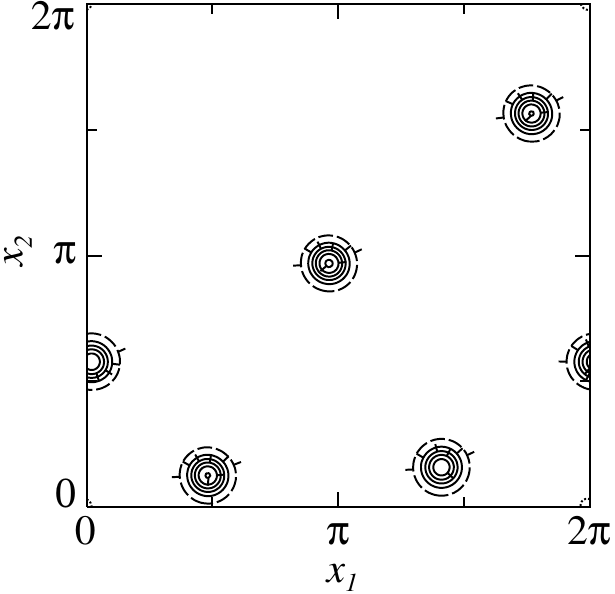}}

~

\centerline{\raisebox{2.3in}{(c)} \ \includegraphics[width=3in]{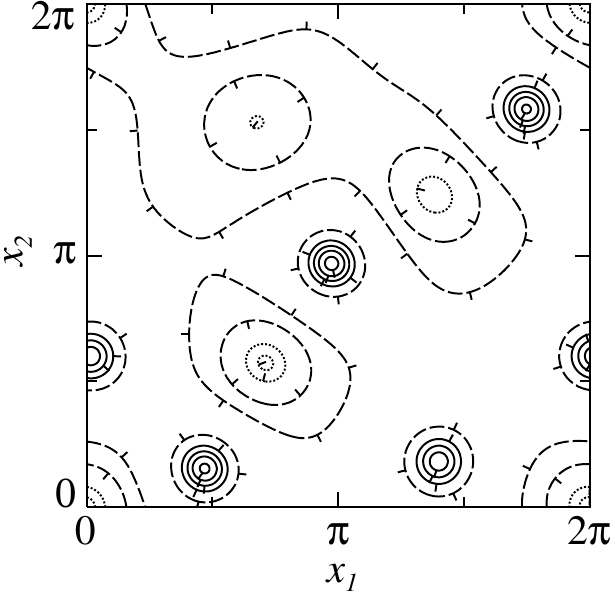}
\hfill\raisebox{2.3in}{(d)} \ \includegraphics[width=3in]{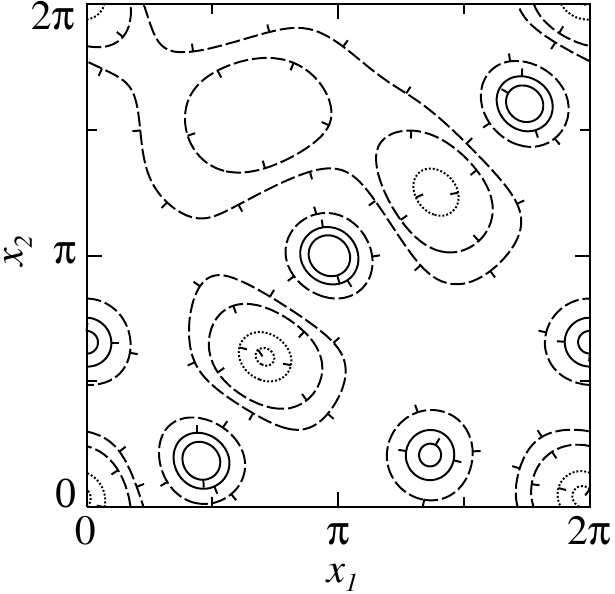}}
\caption{Maximum slow-time growth rates \rf{mgr} of large-scale modes due to
the action of the $\alpha$-effect (vertical axis) versus dominant fast-time growth
rates of small-scale modes (horizontal axis) for $\eta=0.05$ in the collection
of 23 family P flows (a). The topography of the vertical component
for three poloidal flows \rf{pol} is shown in (a) by gray filled circles
marked P$_1$ (b), P$_2$ (c) and P$_3$ (d). Isolines show the values step 2
of the r.m.s.-normalised factor $\nabla^2p(x_1,x_2)$ in the vertical component
of sample flows.
Dotted, dashed and solid lines indicate negative, zero and positive values,
respectively; small ticks point in the direction of decreasing values.}
\label{pigs}\end{figure}

\begin{table}[p]
\caption{Dynamo properties of the three family P flows \rf{pol}
shown in \xf{pigs}a by gray filled circles for $\eta=0.05$.}
\center\begin{tabular}{|c|c|c|c|c|c|}\hline
Flow&$\gamma_{\rm sm}$&$\gamma_\alpha$&$\nu_1$&$\nu_2$&$\nu_3$\\\hline
P$_1$&-0.05066&0.3448$\times10^{-4}$&1.597967&7.905751&8.551576\\
P$_2$&-0.04314&2.083$\times10^{-4}$&-1.708137&3.269258&4.412460\\
P$_3$&-0.03965&4.746$\times10^{-4}$&-1.619310&1.531359&2.172698\\\hline
\end{tabular}\label{tab0}\end{table}

We have computed solutions to the ``nonlinear eigenvalue problem''
\rf{red}, where the r.h.s.~\rf{fi} involves three unknown parameters $\nu_j$:
$$f(g)=\nu_1\e^g+\nu_2\e^{-g}+\nu_3g^3.$$
For the employed resolution of $128^2$ Fourier harmonics,
the energy spectrum decay of the solutions $g(x_1,x_2)$
is in the range 17--21 orders of magnitude. The same vertical profile
$$\widetilde{g}(x_3)=1+\cos x_3+\sin\,2x_3$$
of the flow potential \rf{pot} has been employed to construct a collection
of 23 sample flows. Small-scale dynamo fast-time growth rates and the maximum
slow-time growth rates of large-scale modes generated by the $\alpha$-effect
have been computed for $\eta$=0.05,
for which none of the 23 poloidal flows \rf{pol} generates small-scale magnetic
field (see \xf{pigs}a). For three flows, Figs.~\ref{pigs}b-d show isolines
of the fields $\nabla^2\,g(x_1,x_2)$ normalised so that their r.m.s.~value is 1;
by \rf{pol}, they reflect the topography of vertical components of the flows.
The three flows (marked by gray filled circles in \xf{pigs}a; see
Table~\ref{tab0}) are well separated in the plane (maximum growth rate of
large-scale modes due to the $\alpha$-effect,
dominant growth rate of small-scale modes). The flow represented by the left
circle P$_1$ features the highest contrast in the component structure, but
the part of the fluid volume, occupied by vigorous vertical jets, is small (see
\xf{pigs}b); apparently, this is responsible for the minimum (over the set)
ability of this flow to sustain small-scale magnetic field and to generate
large-scale field by means of the $\alpha$-effect. The intermediate circle
P$_2$ represents a flow featuring a comparable contrast, but there is
a significant increase in the part of the volume where vertical motion is
relatively fast (see \xf{pigs}c). Consequently, for this flow we observe
a decrease in the fast-time decay rate of small-scale field and an increase
in the slow-time growth
rate of the generated large-scale field. Finally, the right circle P$_3$
represents the flow of the least contrast (among the three flows under
discussion), but the part of the volume of relatively fast vertical motion
has again increased (see \xf{pigs}d), accompanied by a further decrease
in the decay rate of small-scale field and an increase in the slow-time growth
rate of large-scale field generated by the $\alpha$-effect.

\subsection{Family V$_1$}\label{alV}

\begin{figure}
\centerline{\includegraphics[width=5in]{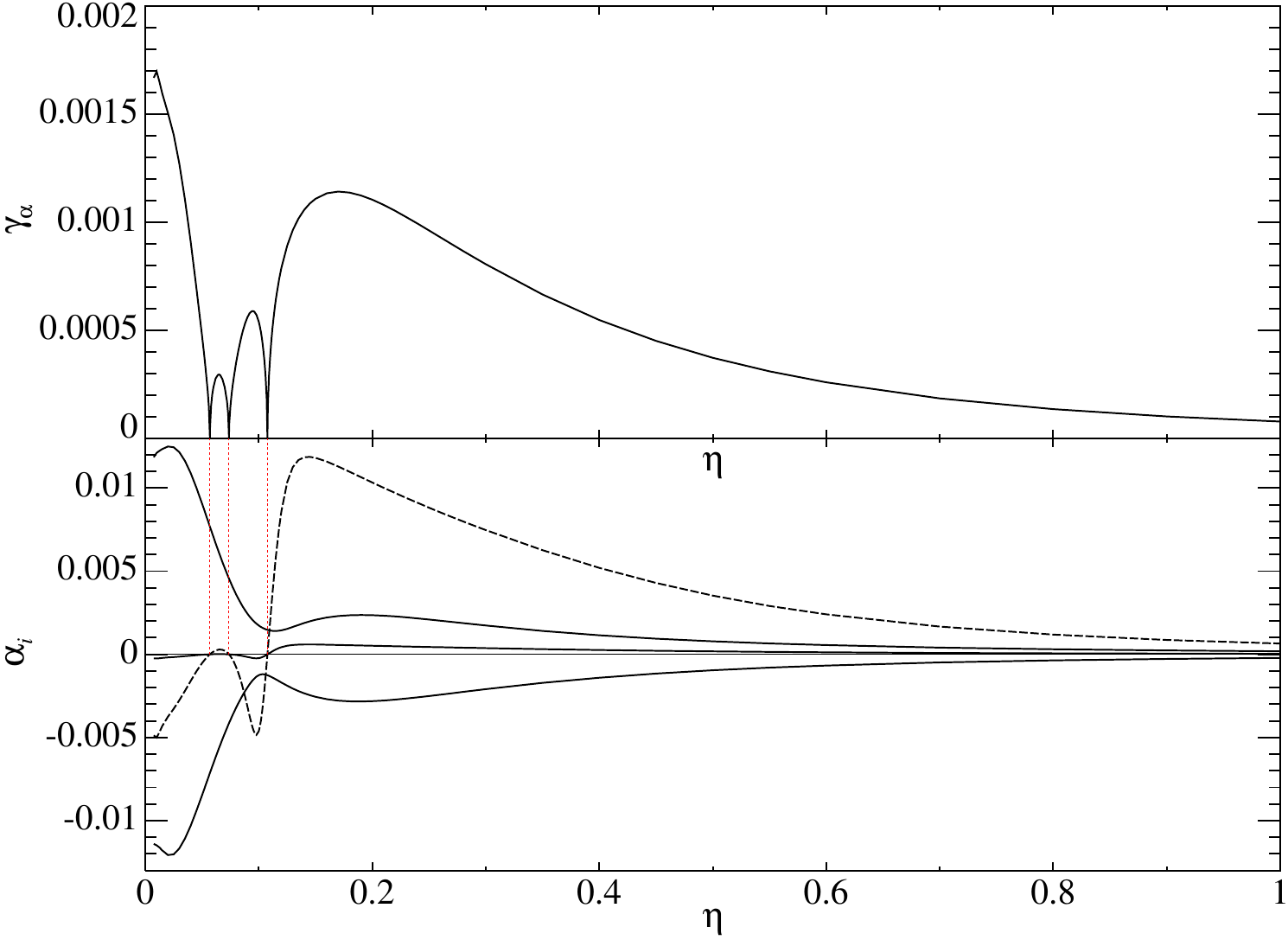}}
\caption{Upper panel: maximum slow-time growth rate of large-scale magnetic field
generated by the $\alpha$-effect (vertical axis) versus magnetic molecular
diffusivity $\eta$ (horizontal axis) for a sample flow from family V$_1$
\rf{V1} constructed for \rf{sfV}.
Lower panel: three eigenvalues $\alpha_i$ of the symmetrised
$\alpha$-tensor $\sA$ for the same flow (solid line) and the intermediate
eigenvalue $\alpha_2$ multiplied by 20 (dashed line). Dotted
vertical lines join the locations of the zero maximum growth rate due to
the $\alpha$-effect with the zeroes of the intermediate eigenvalue of $\sA$.}
\label{rva}\end{figure}

We have computed slow-time growth rates $\gamma_\alpha$ of large-scale magnetic
field generated by the $\alpha$-effect in a sample family V$_1$ flow \rf{V1}
constructed for randomly chosen functions and constants
\be U_1(x_1)&=-2\cos x_1+1.5\sin x_1+0.5\sin 2x_1+0.75\cos 3x_1
-0.2\sin 3x_1-0.1\sin 4x_1,\nonumber\\
U_2(x_2)&=\cos(\e^{\cos 3x_2}-\sin 2x_2)-0.3\sin x_2,\label{sfV}\\
U_3(x_3)&=\sin x_3-0.75\cos 2x_3+0.25\cos 3x_3+0.2\sin 4x_3,\nonumber\\
C_1&=1,\qquad C_2=2,\qquad C_3=-3.\nonumber\end{align}
For this flow, the plot of $\gamma_\alpha$ (see the upper panel of
\xf{rva}) has a rather intricate structure: by virtue of \rf{mgr}, it falls off
to zero each time the intermediate eigenvalue $\alpha_2$ of the symmetrised
$\alpha$-tensor $\sA$ vanishes (in order to visualise legibly these zeroes,
the product $20\alpha_2$ is shown in the lower panel of \xf{rva} by a dashed
line). This is reminiscent of the ``window'' in $\eta$, in which small-scale
generation by the 1:1:1 ABC-flow fails \cite{ArK} --- however, in the present
case intervals where there is no dynamo degenerate into individual points.

\subsection{Family V$_2$}\label{alS}

\begin{figure}[t]
\centerline{\raisebox{2in}{(a)} \ \includegraphics[width=3in,height=2.3in]{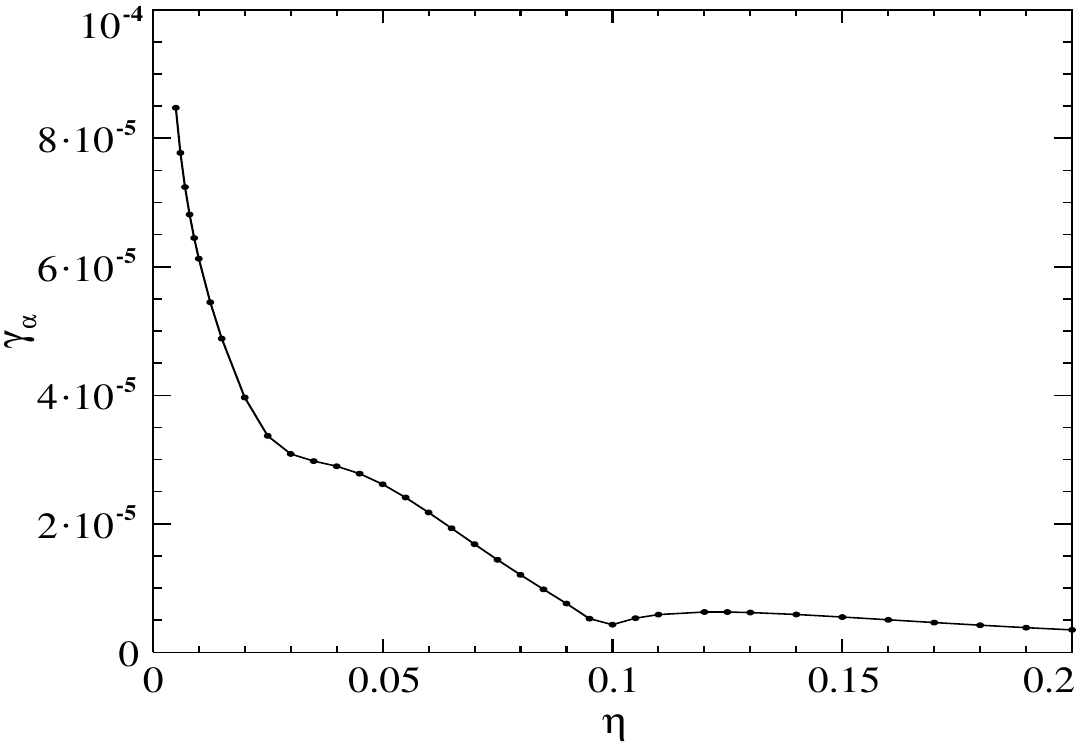}
\hfill\raisebox{2in}{(b)} \ \includegraphics[width=3in,height=2.3in]{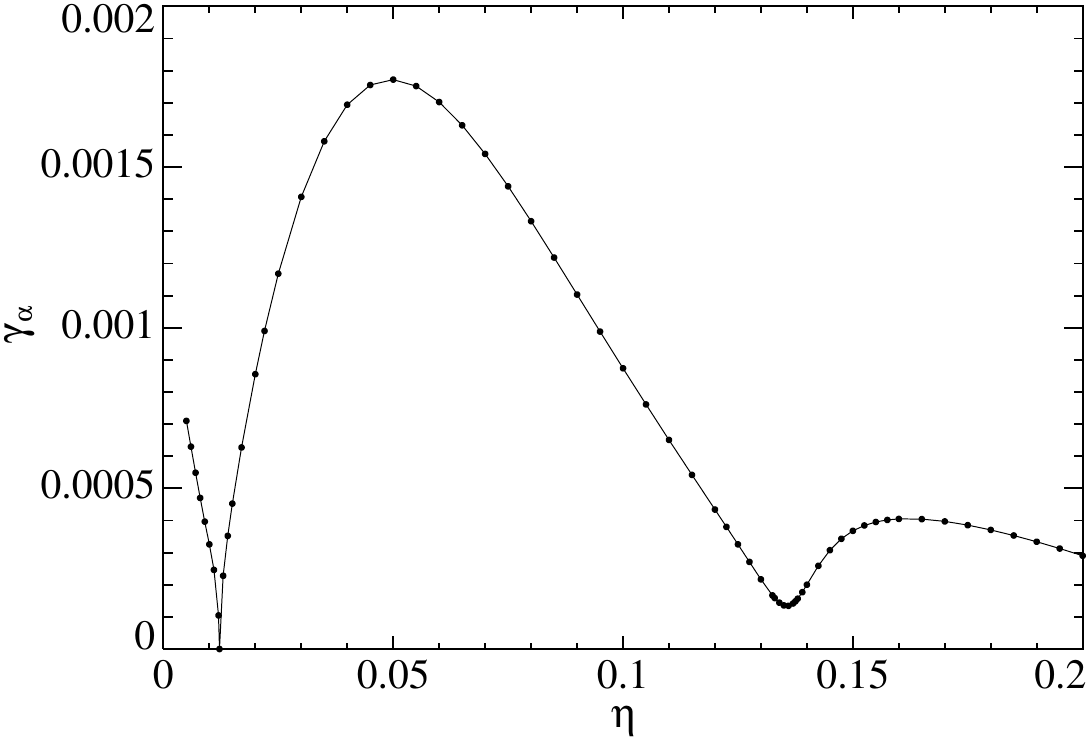}}

~

\centerline{\raisebox{2in}{(c)} \ \includegraphics[width=3in,height=2.3in]{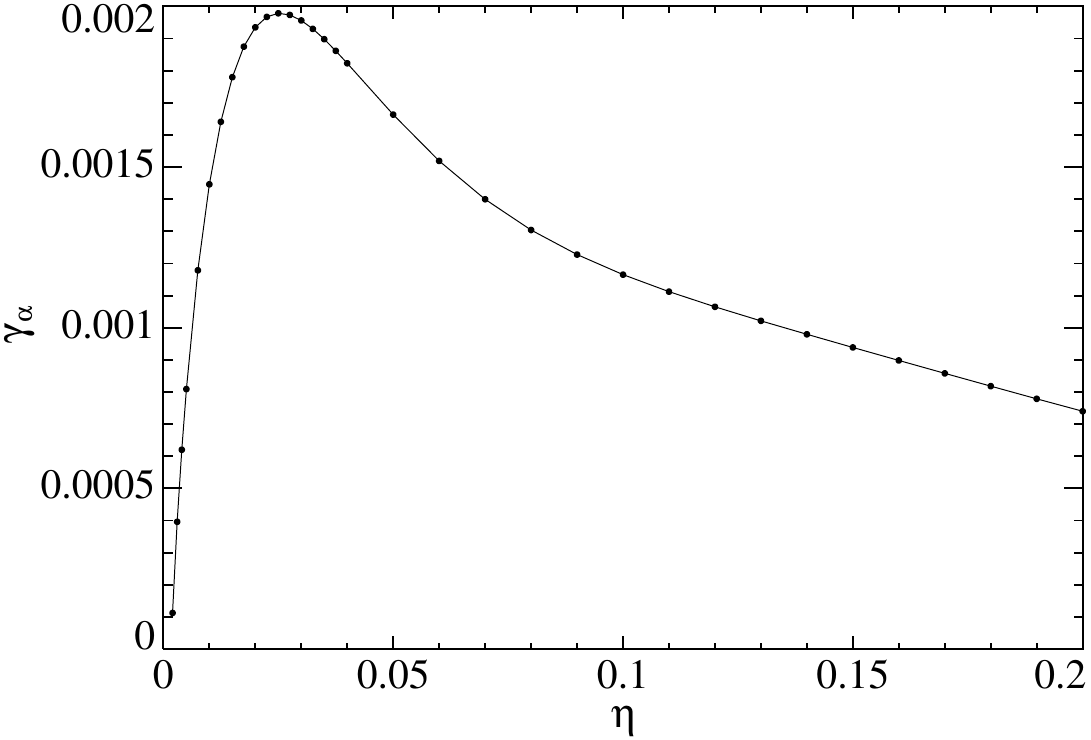}
\hfill\raisebox{2in}{(d)} \ \includegraphics[width=3in,height=2.3in]{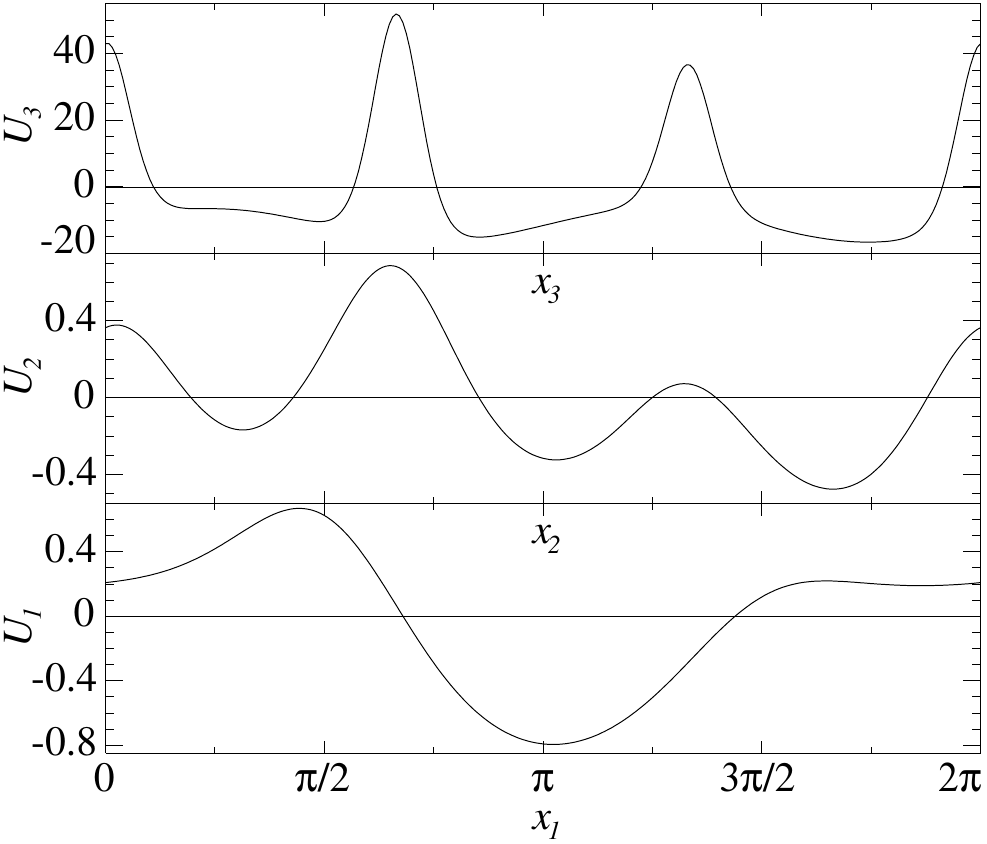}}
\caption{Maximum slow-time growth rate of large-scale magnetic field generated
by the $\alpha$-effect (vertical axis) versus magnetic molecular diffusivity
(horizontal axis) in the three sample flows \rf{V2} from family V$_2$ constructed
for \rf{v2r} (a), the r.m.s.-normalised functions $U_i$ \rf{uu} and \rf{Ci1} (b),
and $U_i$ that are Fourier series with pseudorandomly generated coefficients (c).
Dots show the computed values. Plots of the three functions $U_i$ \rf{uu} (d).}
\label{rsa}\end{figure}

We plot in \xf{rsa}a the maximum slow-time growth rate \rf{mgr} of large-scale magnetic
field generated by the $\alpha$-effect in a sample flow from family V$_2$
\rf{V2} constructed for
\se{v2r}\be u_1(x_1)&=\e^{(\sin x_1)/4-\cos 2x_1}+(\cos x_1)/2,\label{u1}\\
u_2(x_2)&=(\e^{(\sin x_2)/4+(\cos 3x_2)/3}+(\cos 2x_2)/6)^2,\label{u2}\\
u_3(x_3)&=\e^{\sin x_3+4\cos 3x_3}+5\sin 2x_3,\label{u3}\\
U_i(x_i)&=u_i(x_i)-\LA u_i(x_i)\RA,\label{uu}\\
C_1&=3,\qquad C_2=2,\qquad C_3=1.\label{uC}\end{align}\end{subequations}
For the smallest molecular diffusivity $\eta=0.005$ considered for this flow,
the energy spectrum of solutions to auxiliary problems of type I, ${\bf S}_1$,
${\bf S}_2$ and ${\bf S}_3$, decays by 5, 3 and 6 orders of magnitude in runs
with the resolution of $128^3$ Fourier harmonics, respectively, and by 9, 7 and
9 orders in $256^3$ harmonics runs. Nevertheless, the discrepancy
in the elements of the $\alpha$-tensor $\A$ and the maximum slow-time growth
rate of large-scale field is below
$2\times10^{-8}$ (owing to the fast energy spectrum decay of the flow).

The growth rates $\gamma_\alpha$ for this flow
are rather small (see \xf{rsa}a). This can be attributed
to its strong anisotropy: while the amplitude of the functions $U_1$ and $U_2$
is below unity, the amplitude of $U_3$ is roughly 60 (see \xf{rsa}d);
the resulting flow \rf{V2} is thus close to a plane-parallel horizontal flow
that is incapable of dynamo action by the Zeldovich \cite{Zel} antidynamo
theorem. This has prompted us to also consider the flow \rf{V2} constructed for all
\BE C_i=1\EE{Ci1}
and the functions $U_i$ \rf{u1}--\rf{uu} used previously but normalised so that
their r.m.s.~value becomes~1. The maximum slow-time growth rates $\gamma_\alpha$
are about an order of magnitude higher (see \xf{rsa}b).

Slow-time growth rates of large-scale magnetic field generated by the $\alpha$-effect
in the flow \rf{V2}, \rf{v2r} are also significantly smaller than those
obtained for a yet another sample family V$_2$ flow (see~\xf{rsa}c).
It involves zero-mean functions $U_i$ that are Fourier series
with pseudorandom coefficients for wave numbers up to 63, whose energy
spectrum decays exponentially by more than 10 orders of magnitude. We have
checked that no small-scale dynamo operates for the considered molecular
diffusivities (by computing the dominant fast-time growth rates
$\gamma_{\rm sm}$ of small-scale modes
for $\eta=0.005$, 0.01, 0.02 and 0.05). A different (compared to the sample
flows discussed above) behaviour of $\gamma_\alpha$ is observed
on decreasing molecular diffusivity; however, the considered values of $\eta$
are still too high to conjecture that the saturated regime for $\eta\to0$
has set in.

\subsection{Family L}\label{alL}

\begin{figure}[t]
\centerline{\raisebox{2in}{(a)}~\includegraphics[width=3in]{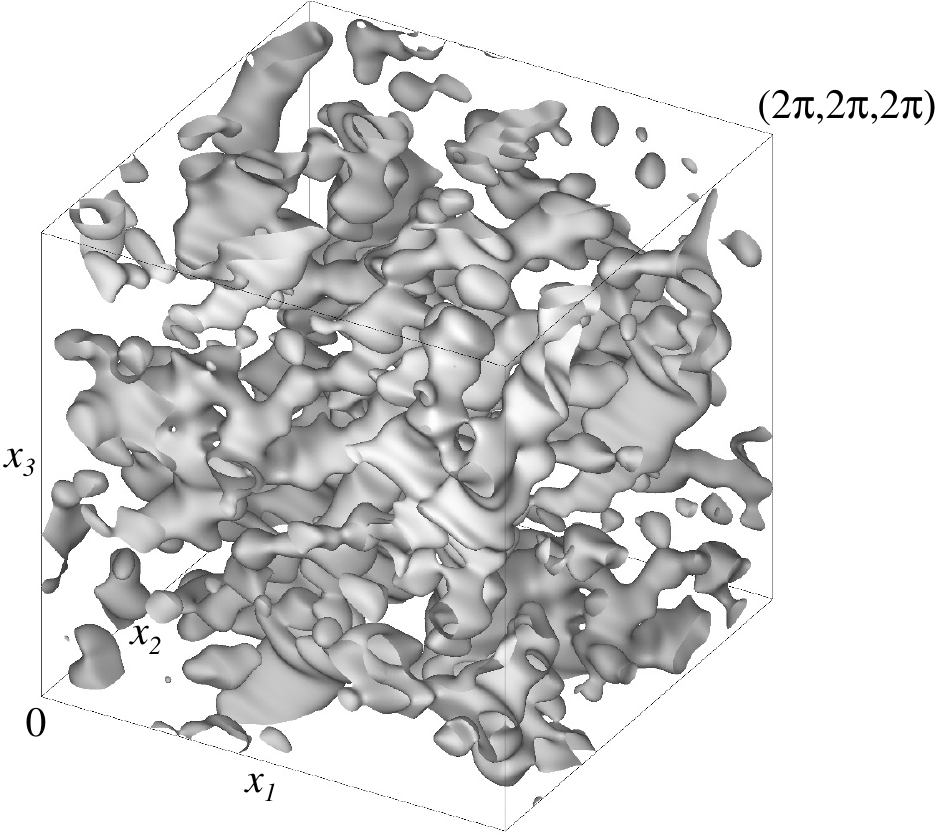}
\hspace*{1em}\raisebox{2in}{(b)}~\includegraphics[width=3in]{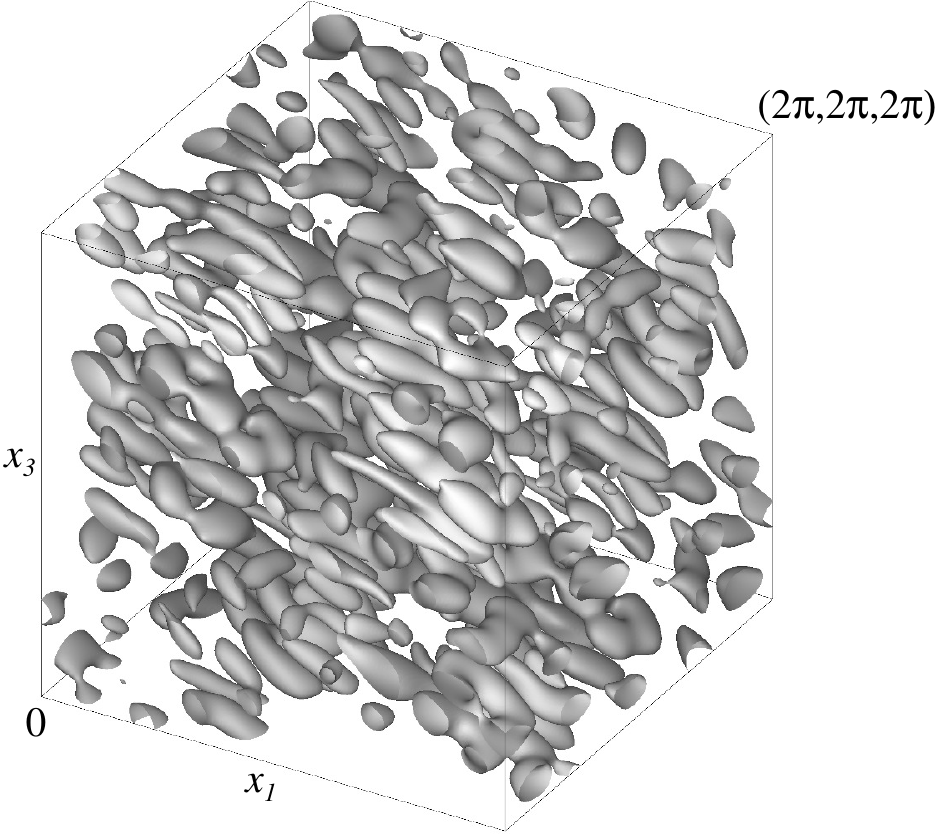}}

~
\centerline{\raisebox{2in}{(c)}~\includegraphics[width=3in]{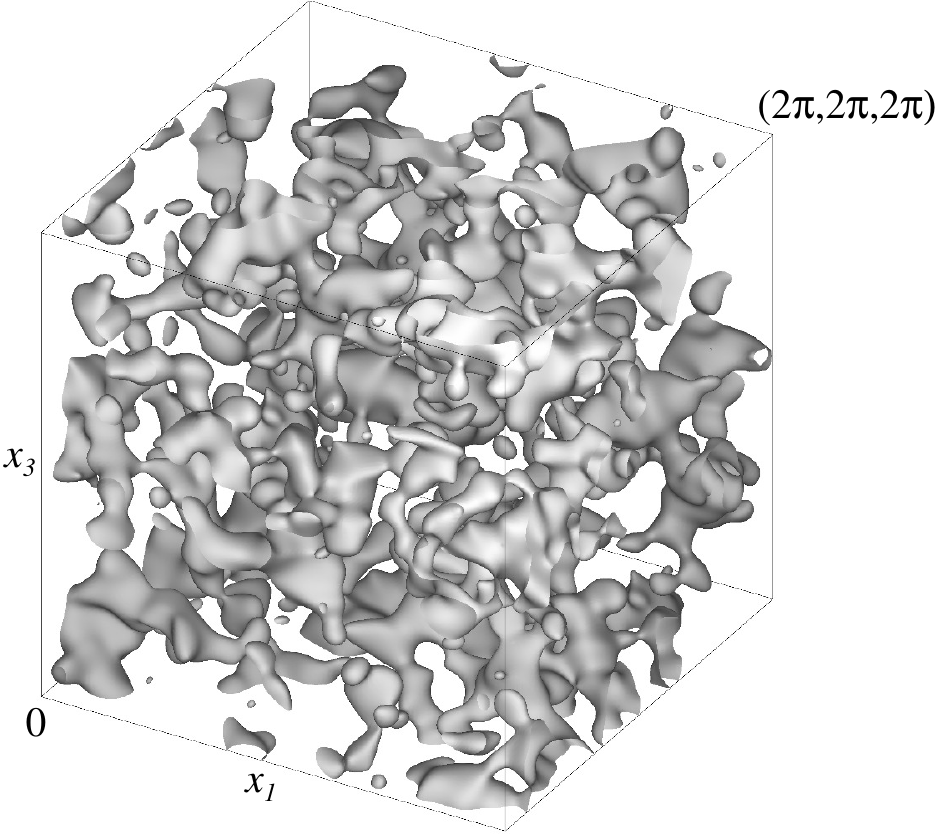}
\hspace*{1em}\raisebox{2in}{(d)}~\includegraphics[width=3in]{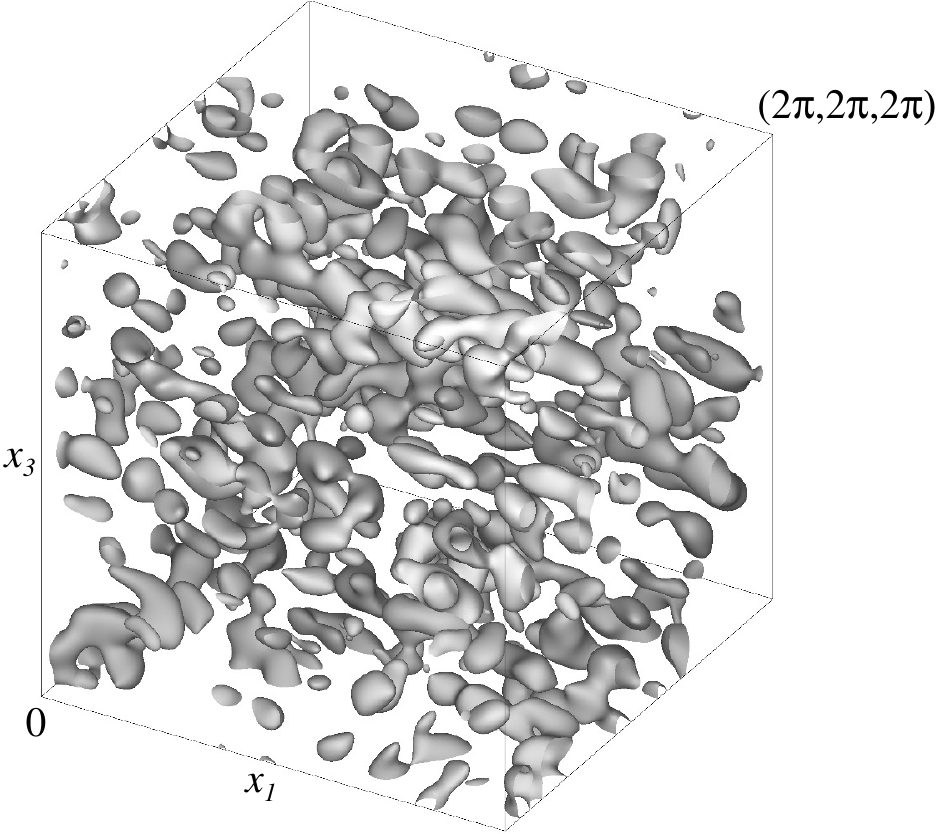}}
\caption{Isosurfaces of the flow velocity (a,c) and vorticity (b,d)
at the levels of 1/3 of the respective maxima for the two sample family L flows
\rf{v2} featuring non-zero helicity spectrum, for which the magnetic $\alpha$-effect
has been explored, and small-scale magnetic field generation starts in the even
(a,b) or odd (c,d) invariant subspace. One periodicity cube $\T^3$ is shown.}
\label{vLa}\end{figure}

\begin{figure}[p]
\centerline{\raisebox{1.8in}{(a)}\includegraphics[width=3in,height=2.1in]{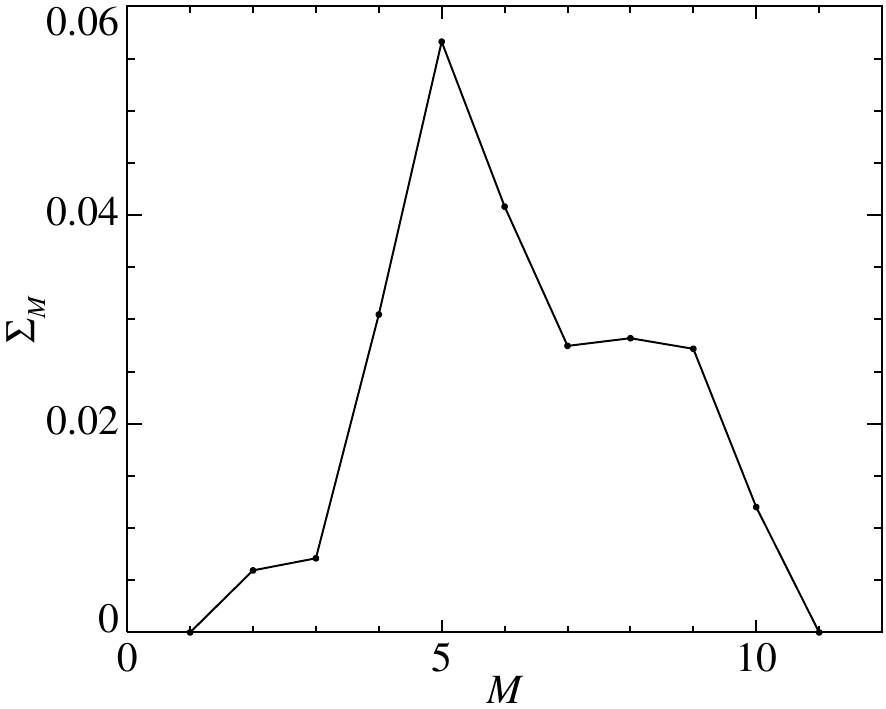}
\hfill\raisebox{1.8in}{(b)}\includegraphics[width=3in,height=2.1in]{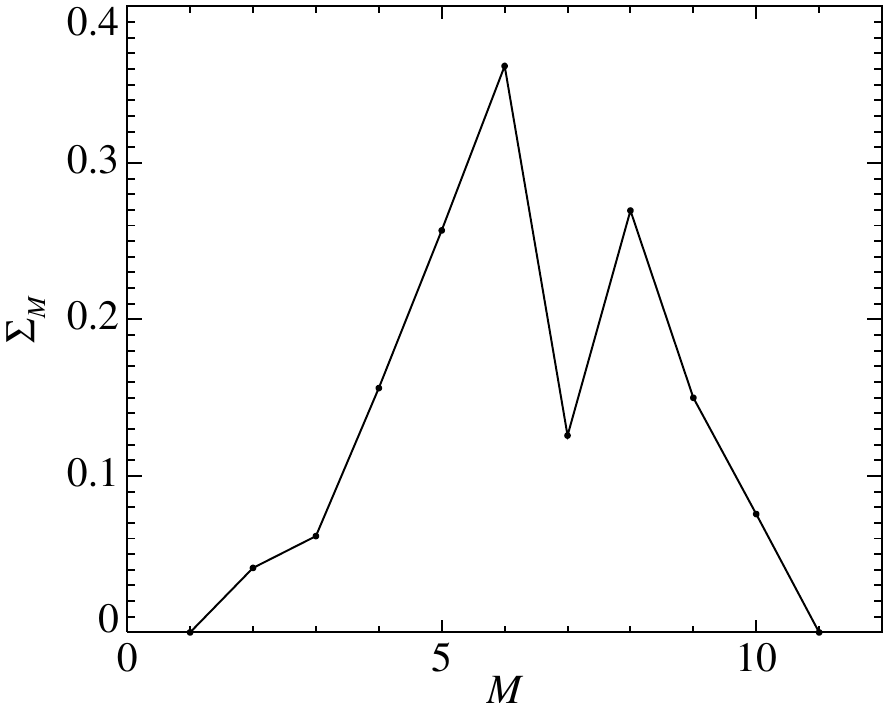}}
\caption{Helicity spectrum seminorm (vertical axis) versus the wave vector
shell number (horizontal axis) for the two sample family L flows under
consideration, in which small-scale magnetic field generation starts
in the even (a) or odd (b) invariant subspace.}
\label{sLa}\end{figure}

\begin{figure}
\centerline{\raisebox{1.8in}{(a)}\includegraphics[width=3in,height=2.1in]{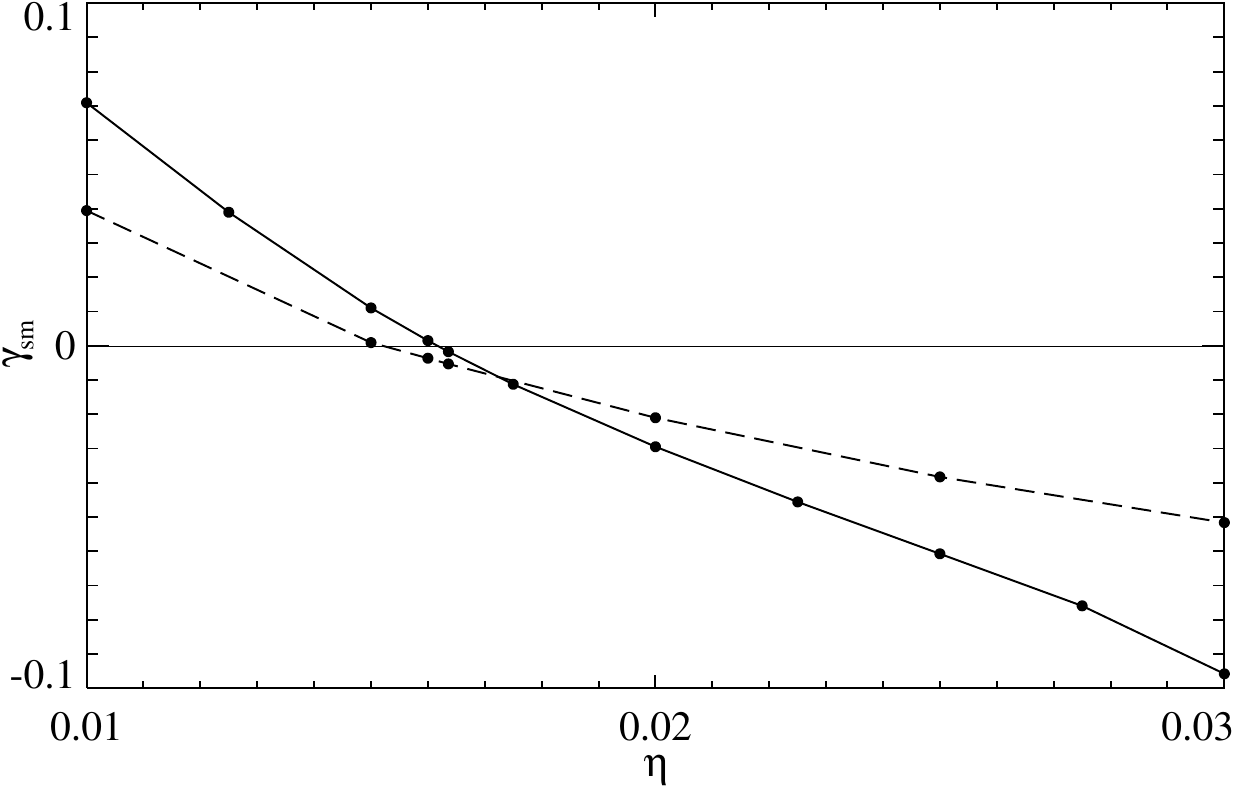}
\hfill\raisebox{1.8in}{(b)}\includegraphics[width=3in,height=2.1in]{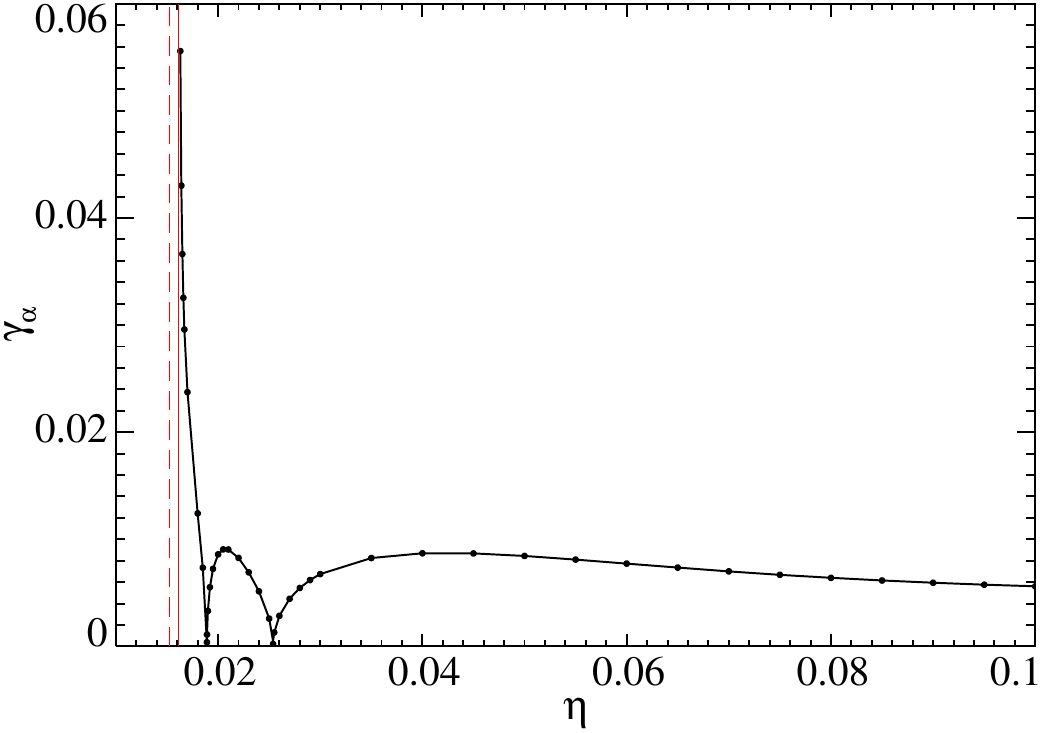}}

~

\centerline{\raisebox{1.8in}{(c)}\includegraphics[width=3in,height=2.1in]{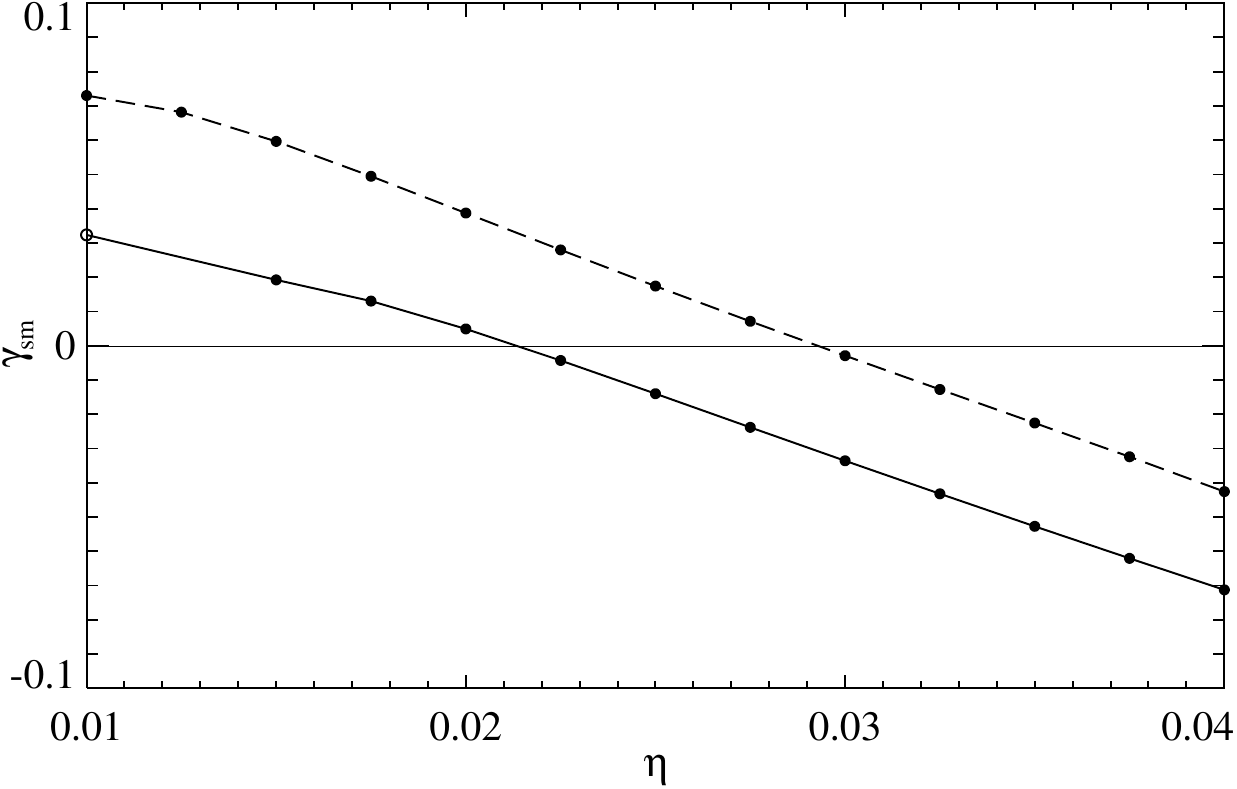}
\hfill\raisebox{1.8in}{(d)}\includegraphics[width=3in,height=2.1in]{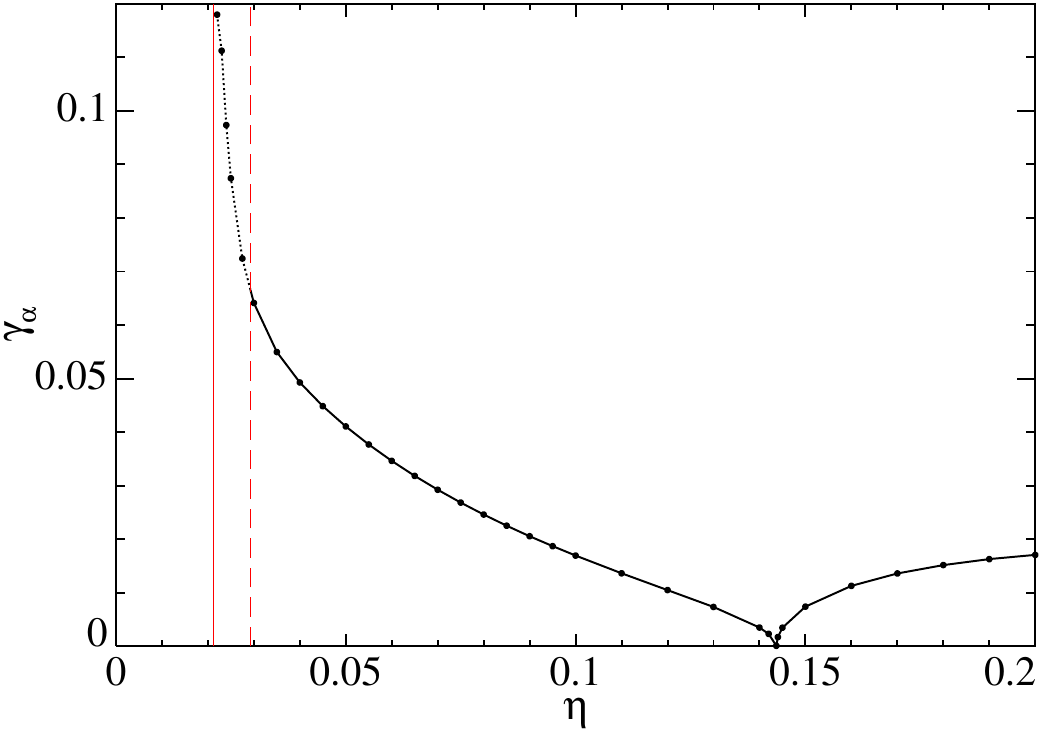}}
\caption{The fast-time growth rate (a,c) of small-scale magnetic modes (vertical
axis) dominant in the even (solid line) and odd (dashed line) invariant
subspaces, and the maximum slow-time growth rate $\gamma_\alpha$ \rf{mgr}
of large-scale magnetic field (b,d) generated by the $\alpha$-effect
(vertical axis) versus magnetic molecular diffusivity (horizontal axes)
for the two sample family L flows under consideration, where
small-scale magnetic field generation starts in the even (a,b) or odd (c,d)
invariant subspace. Dots show the computed values. In panels (b,d), thin
vertical lines are located at the critical molecular diffusivities
for the onset of the small-scale dynamo in the even (solid lines) or odd
(dashed lines) subspaces. In (d), dotted line shows the part of the graph
of the slow-time growth rate for $\eta$ between the critical values
for the onset of small-scale generation in the two subspaces, for which
large-scale generation by the $\alpha$-effect
is overshadowed by the small-scale one in the odd subspace.}
\label{gLa}\end{figure}

Two sample family L flows \rf{v2} have been considered for the Monge potentials
$A$ and $B$ that are eigenfunctions of the Laplace operator associated
with the eigenvalue $-26$; they are sums of 72 Fourier harmonics, whose wave
vectors are composed of numbers $\pm3,\pm4,\pm1$ or of $\pm5,0,\pm1$, in both
sets in any order and for any combination of signs. The harmonics enter
the sums with complex pseudorandom coefficients (complex conjugacy is enforced
for the flows to be real). Isosurfaces of the velocity $|\bf v|$ and vorticity
$|\nabla\times\bf v|$ of these flows in \xf{vLa} illustrate their strong
spatial intermittency; \xf{sLa} showing helicity spectrum seminorms
$$\Sigma_M=\sum_{M-1<|{\bf m}|\le M}|H_{\bf m}|$$
furnishes evidence that the helicity spectra of the two flows do not vanish.
The plots of the maximum slow-time growth rate \rf{mgr} of large-scale
magnetic field generated by the $\alpha$-effect, as functions of magnetic
molecular diffusivity (see \xf{gLa}b,d), involve square-root type cusps where
the growth rates vanish together with the intermediate eigenvalue $\alpha_2$
of the symmetrised $\alpha$-tensor $\sA$; this is the same phenomenon as
the one seen in \xf{rva}.

Notably, in \xf{gLa}b,d the graphs of the maximum slow-time growth rate
$\gamma_\alpha$ have vertical asymptotes.
A similar singular behaviour of eddy diffusivity was encountered
(see section~3.7 in \cite{VZ}), but to the best of our knowledge it was never
documented for the $\alpha$-effect. The nature of this phenomenon is the same
for both mechanisms of large-scale generation, as we will now briefly discuss.
Let us consider a first auxiliary problem \rf{Seq} in the form of the left
equation for a zero-mean solenoidal field ${\bf S}_k$. Generically,
the magnetic induction operator $\L$ has a three-dimensional kernel spanned
by the fields ${\bf e}_k+{\bf S}_k$, thus being invertible in the functional
subspace of our interest. We expand, for a given $\eta$, the unknown field
and the r.h.s.~of the left equation \rf{Seq} in the basis of solenoidal
zero-mean eigenfunctions ${\bf f}_n(\eta)$ of the magnetic induction operator,
$\L{\bf f}_n(\eta)=\mu_n(\eta){\bf f}_n(\eta)$:
$${\bf S}_k(\eta)=\sum_n\sigma_{nk}(\eta){\bf f}_n(\eta),\qquad
-{\partial{\bf v}\over\partial x_k}=\sum_n\alpha_{nk}(\eta){\bf f}_n(\eta).$$
(For the sake of argument, we assume that $\L$ does not involve Jordan form
cells of size 2 or more, although it is not difficult to take them into
account in a fully formal proof.) Then, evidently,
\BE\sigma_{nk}(\eta)={\alpha_{nk}(\eta)\over\mu_n(\eta)}\quad\Rightarrow\quad
{\bf S}_k(\eta)=\sum_n{\alpha_{nk}(\eta)\over\mu_n(\eta)}{\bf f}_n(\eta);\EE{Sso}
while $\L$ is invertible, all $\mu_n(\eta)\ne0$ and $\mu_n(\eta)\to-\infty$,
the series thus remaining well-defined and convergent.

Suppose now $\eta\to\eta_{\rm cr}$ for the onset of the small-scale magnetic
field generation, i.e., $\mu_N(\eta_{\rm cr})=0$ for some $N$ (corresponding
in our case to the dominant mode; again, to simplify the argument, we assume
that the emerging eigenvalue zero has multiplicity one; it is important that
the eigenvalue at the onset is real). The eigenfunction ${\bf f}_N(\eta)$
remains smooth and bounded, however, the solution \rf{Sso} infinitely increases,
as well as the $\alpha$-effect tensor
$$\A_k(\eta)={\alpha_{Nk}(\eta_{\rm cr})\over\mu_N(\eta)}\LA{\bf v}\times
{\bf f}_N(\eta_{\rm cr})\RA+{\rm o}(\mu_N^{-1}).$$
This results in emergence of the vertical
asymptote like the one shown in \xf{gLa}b. (Note that this may be
interpreted not as ``an infinite rate of generation'' at $\eta=\eta_{\rm cr}$,
but rather as that the ansatz \rf{ble} becomes inapplicable for the critical
$\eta$. Although $|\lambda_1|\to\infty$ when $\eta_{\rm cr}$ is approached,
Re\,$\lambda_1$ approximates the slow-time growth rate only for $\varepsilon\to0$;
for acceptable $\varepsilon$, the product $\varepsilon{\rm Re}\lambda_1$
approximating the fast-time growth rate remains finite, if does not tend to zero.)

There is a subtlety: the Monge potentials $A$ and $B$ defining a family L flow
are linear combinations of Fourier harmonics that are eigenfunctions
of the Laplacian associated with the same eigenvalue. Because
of the periodicity condition, a wave vector of each harmonics has integer
components, whose sum has the same parity as the Laplacian eigenvalue.
Therefore, any flow \rf{v2} belongs to what we call the ``even'' subspace
composed of harmonics such that the sum of wave numbers is even.
In the ``odd'' complementary subspace, the sum of wave numbers of
the constituting harmonics is odd. When the flow belongs to the even subspace,
even and odd subspaces are invariant for the small-scale magnetic induction
operator $\L$. All neutral modes ${\bf e}_k+{\bf S}_k$ belong to the even
subspace and do not ``feel'' the onset of small-scale field generation, when
it occurs (on decreasing $\eta$) in the odd subspace. The $\alpha$-effect tensor
acquires singularity, as discussed above, when the onset is in the even
subspace. This happens (see \xf{gLa}a,b) for the sample flow shown in \xf{vLa}a,b.
For the flow shown in \xf{vLa}c,d, the onset occurs in the odd subspace (see
\xf{gLa}c) not affecting the graph of the maximum slow-time growth rate
of large-scale magnetic field generated by the $\alpha$-effect \xf{gLa}d.
(For this flow, the part of the graph of $\gamma_\alpha$ shown in \xf{gLa}d
by the dotted line for $\eta$ between the critical values for the onset
of small-scale generation in the two subspaces only illustrates the singular
behaviour near the critical molecular
diffusivity in the even subspace. As discussed in the introduction
to the present section, large-scale generation, whose fast-time growth rate
is order $\varepsilon$, is overshadowed in this interval by the small-scale
one in the odd subspace, whose fast-time growth rate is order unity.)
Actually, small-scale magnetic field generation occurs in the odd subspace
more frequently than in the even one. Perhaps, the reason is in
that the shortest wave vector in the even subspace is $\sqrt2$ times
longer than the one in the odd subspace, and hence the small wave number
harmonics (storing a significant part of energy of a dominant small-scale
magnetic mode) are more amenable to molecular diffusion in the even subspace
than in the odd~one.

\section{Magnetic eddy diffusivity in non-helical flows: numerical results}\label{ed}

We have examined numerically magnetic eddy diffusivity of some instances
of flows belonging
to families L, C and V$_1$ (see subsections \ref{fL}, \ref{fC} and \ref{fV},
respectively), that are parity-invariant and hence lack the $\alpha$-effect.
Family V$_2$ flows are not parity-invariant (see a comment to this effect
at the end of section~\ref{fV}) and generically sustain the $\alpha$-effect;
hence, they have been excluded here from examination. As when exploring
the $\alpha$-effect, for each pair (a flow / magnetic molecular diffusivity)
considered here, we have computed the fast-time growth rate $\gamma_{\rm sm}$ of the dominant
small-scale magnetic mode, and in what follows we comment on the minimum
magnetic eddy diffusivity \rf{miE} for flows that do not generate small-scale
magnetic field. The rationale is discussed in the introduction to section
\ref{al}.

Each employed flow has a unit r.m.s.~velocity.

Computation of the eddy diffusivity tensor
can be significantly simplified (to the extent that there may be no need
to solve auxiliary problems for the adjoint operator) in the presence
of certain symmetries of the flow. In particular, the symmetries of the cosine
flows imply splitting of the domain of the magnetic induction operator
into invariant subspaces and a special structure of the eddy diffusivity tensor.
We discuss these issues in section~\ref{edC}.

\subsection{Family L}\label{edL}

\begin{figure}[p]
\centerline{\raisebox{2.4in}{(a)}~\includegraphics[width=3in]{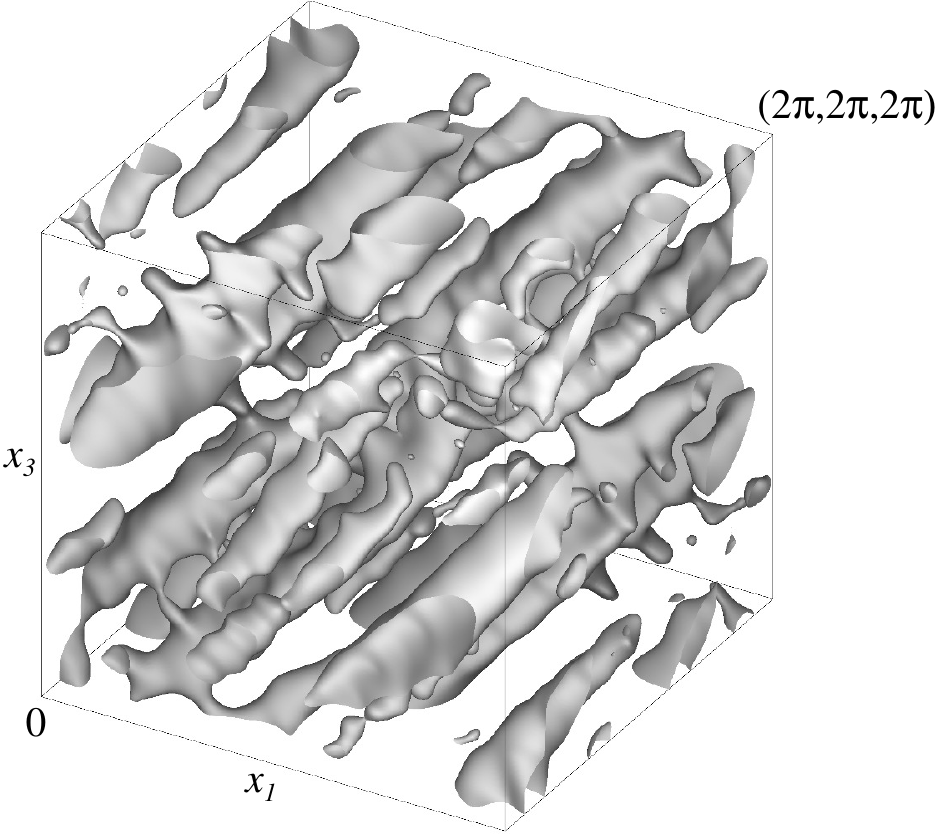}
\hspace*{2em}\raisebox{2.4in}{(b)}~\includegraphics[width=3in]{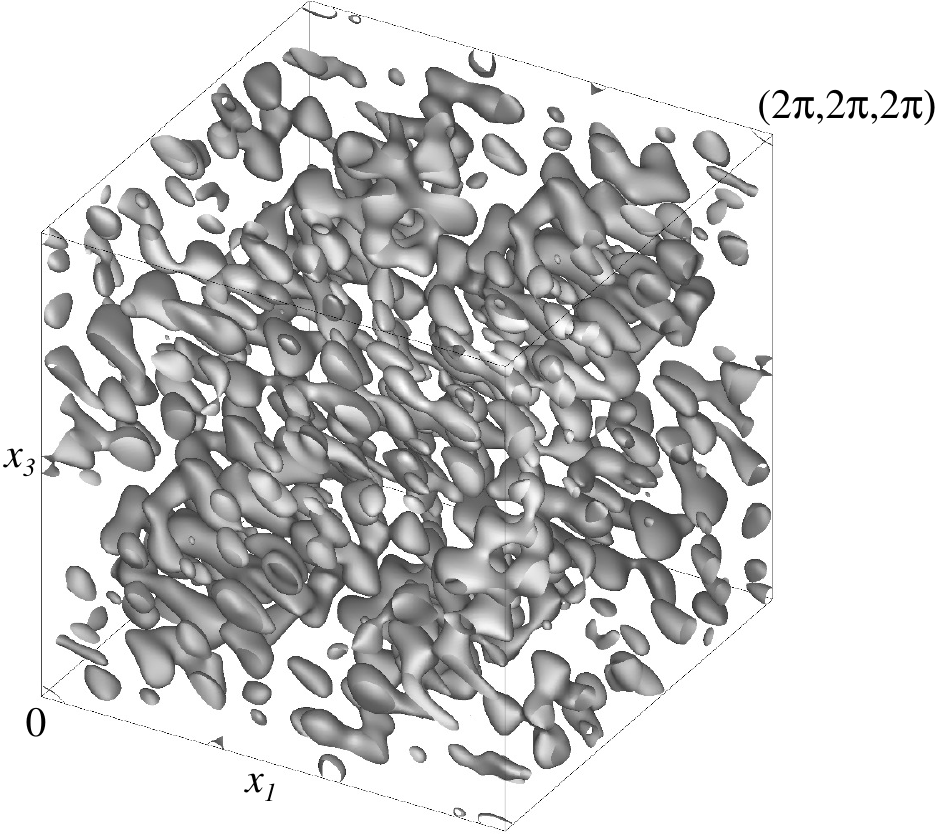}}
\caption{Isosurfaces of the flow velocity (a) and vorticity (b) at the levels
of 1/3 and 2/5 of the maxima, respectively, for a sample parity-invariant
family L flow \rf{v2} for which magnetic eddy diffusivity has been explored.
One periodicity cube $\T^3$ is shown.}
\label{vL}\end{figure}

\begin{figure}[p]
\centerline{\includegraphics[width=7in,height=4in]{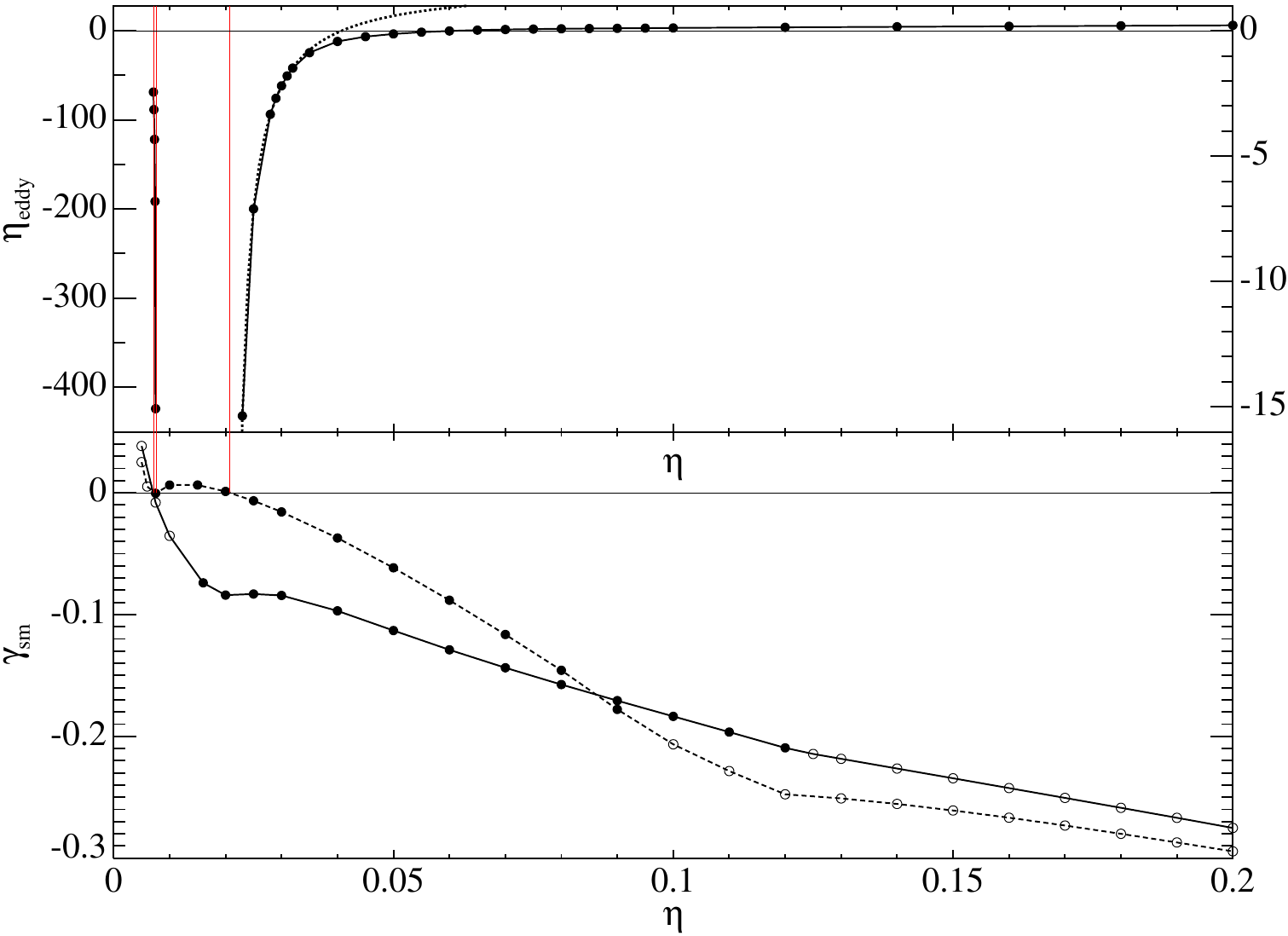}}
\caption{Upper panel: minimum magnetic eddy diffusivity $\eta_{\rm eddy}$
(vertical axes) versus molecular diffusivity $\eta$ (horizontal axis)
in the sample parity-invariant family L flow \rf{v2}. Dots show the computed
values. The scales for the left and right branches of the plot are shown
in the left and right vertical axes, respectively. Dotted curve is
the least-squares hyperbolic fit for the 7 shown
values of $\eta_{\rm eddy}$ computed in the interval $0.023\le\eta\le0.032$~.
Lower panel: dominant fast-time growth rate of small-scale modes (vertical
axis) as a function of $\eta$ for the sample flow in the invariant subspaces
of parity-invariant (solid line) and parity-antiinvariant (dashed line) vector
fields. Filled (hollow) circles indicate that eigenvalues of the small-scale
magnetic induction operator are real (complex, respectively). Thin vertical
lines are located at the critical molecular diffusivities for the onset
of the small-scale dynamo in the two symmetry subspaces.}
\label{eL}\end{figure}

Like in section \ref{alL}, we consider here a sample family L flow \rf{v2}
for the Monge potentials $A$ and $B$ that are eigenfunctions of the Laplace
operator associated with the eigenvalue $-26$. They are again linear
combinations of Fourier harmonics for wave vectors $(\pm3,\pm4,\pm1)$ and
permutations, as well as $(\pm5,0,\pm1)$ and permutations. However,
the pseudorandom coefficients in the combinations are now imaginary, whereby
each potential is an odd (i.e., $f(-{\bf x})=-f({\bf x})$) real scalar field,
resulting in parity invariance of the flow. Isosurfaces of the flow velocity
and vorticity (see \xf{vL}) reveal its elaborate internal structure.

The upper panel of \xf{eL} presents the minimum magnetic eddy diffusivity
\rf{miE} for this flow as a function of the magnetic molecular eddy
diffusivity $\eta$. For $\eta$ sufficiently large,
$\eta_{\rm eddy}>0$, but on decreasing molecular diffusivity $\eta_{\rm eddy}$
changes the sign near $\eta\approx0.0612$. The plot of $\eta_{\rm eddy}$ has
a vertical asymptote located at the critical molecular diffusivity
$\eta_{\rm cr}\approx0.0207$ for the onset of the small-scale magnetic field generation
(see the lower panel of \xf{eL}). Section 3.7 of \cite{VZ} explains this
phenomenon: solutions to the auxiliary problems involve the inverse
small-scale magnetic induction operator $\L^{-1}$; in short,
when $\eta$ is close to $\eta_{\rm cr}$, the norm of $\L^{-1}$ is large, and thus
either solutions to auxiliary problems of type II are large (if the neutral
zero-mean magnetic mode emerging at $\eta=\eta_{\rm cr}$ is parity-invariant),
or solutions to all auxiliary problems are large (if the mode is
parity-antiinvariant). As an illustration, we have also plotted by a dotted
line the least-square hyperbolic fit (the ratio of two linear functions)
obtained for the seven computed values of $\eta_{\rm eddy}$ in the interval
$0.023\le\eta\le0.032$~. The position of the vertical asymptote of the fit,
at the zero of the denominator, differs from the linearly estimated critical
molecular diffusivity $\eta_c\approx0.0207$ by less then $10^{-4}$.

Small-scale parity-antiinvariant magnetic field is generated
for $0.0077\ls\eta\ls0.0207$. For smaller $\eta$ in the adjacent
interval $0.0063\ls\eta\ls0.0077$, generation in this symmetry
subspace ceases. This is analogous to the window where the 1:1:1 ABC-flow
does not act as a small-scale kinematic dynamo \cite{ArK}. Inside this
interval, at $\eta\approx0.0064$, the branch of dominant parity-antiinvariant
magnetic modes is replaced by a different one:
the associated eigenvalues change from real (for large $\eta$) to complex ones
(for small $\eta$), the imaginary part of the dominant eigenvalue experiencing
a jump. In the same interval, at $\eta\approx0.0071$, generation of small-scale
parity-invariant field sets in. (All the critical values have been determined
by linear interpolation of real parts of the computed eigenvalues.)

Thus, no small-scale generation takes place in a short interval
$0.0071\ls\eta\ls0.0077$~. At the right endpoint of this interval,
$\eta\approx0.0077$, there exists a neutral (i.e., the associated eigenvalue
of the magnetic induction operator is 0) parity-antiinvariant zero-mean
small-scale magnetic mode. By the same argument as for $\eta_c\approx0.0207$,
the endpoint hosts another singularity of $\eta_{\rm eddy}$. The minimum eddy
diffusivity tends to $-\infty$ on increasing $\eta$ towards the right
endpoint $\eta\approx0.0077$ (see the left branch of the plot in the upper
panel of \xf{eL}). Thus, we have encountered an example of a flow that has
at least two windows in magnetic molecular diffusivity where no small-scale
magnetic field is generated, but generation of large-scale field does take
place. At $\eta\approx0.0063$, the increment of magnetic field also vanishes,
but the respective eigenvalue of the magnetic induction operator has a non-zero
imaginary part, and consequently $\eta_{\rm eddy}$ remains non-singular.

\subsection{Cosine flows: family C}\label{edC}

Numerical investigation of eddy diffusivity of the cosine flows \rf{cos} is
significantly simplified by their two properties: translation by half the period
in the vertical direction reverses the flows, and they are symmetric in $x_3$.

If translation by a vector $\bf a$ reverses the flow,
\BE\bf v(x+a)=-v(x)\EE{tr}
($|\bf a|$ is then a half of the smallest period in the direction of $\bf a$),
then by virtue of \rf{Seq} and \rf{Sml}
\BE{\bf S}^-_l({\bf x})={\bf S}_l({\bf x+a})\EE{Sl}
for all $l$. Thus, for such flows it suffices to solve the 3 auxiliary
problems \rf{Seq} of type I. For the cosine flows \rf{cos} this happens for
${\bf a}=(\pi/n)\,{\bf e}_3$.

Furthermore, following \cite{ABNZ}, we use \rf{tr}, \rf{Zl} and \rf{Sl}
for the index $k$ instead of $l$ to transform \rf{Dlmk} into
$$\D^l_{mk}=\eta\int_{\T^3}{\bf Z}_l({\bf x})\cdot\left
(2\,\nabla\times{\partial\over\partial x_m}{\bf Z}_k({\bf x+a})
-{\bf e}_m\times\nabla^2{\bf Z}_k({\bf x+a})\right)\,{\d{\bf x}\over(2\pi)^3}.$$
Integrating here by parts the first scalar product and exploiting
self-adjointness of the Laplacian and the curl, as well as the antisymmetry of
the triple product with respect to permutation of its factors, we find that
$$\De^l_{mk}=-\De^k_{ml}\qquad\Rightarrow\qquad\De^k_{mk}=0$$
for all indices $l,m,k$. Therefore, $\sD=0$ implying $d=0$, and
for any wave vector $\bf q$ eigenvalues \rf{dei} of the eddy diffusivity
operator are two-fold.

A field ${\bf f}=(f^1,f^2,f^3)$ is called
{\it symmetric in $x_i$}, if for all $i$ and $j$
$$f^j((-1)^{\delta^1_i}x_1,(-1)^{\delta^2_i}x_2,(-1)^{\delta^3_i}x_3)
=(-1)^{\delta^j_i}\,f^j({\bf x})$$
and {\it antisymmetric in $x_i$}, if for all $i$ and $j$
$$f^j((-1)^{\delta^1_i}x_1,(-1)^{\delta^2_i}x_2,(-1)^{\delta^3_i}x_3)
=(-1)^{1-\delta^j_i}\,f^j({\bf x}).$$
The symmetry in $x_i$
of the flow $\bf v$ implies that vector fields symmetric or antisymmetric
in $x_i$ constitute invariant subspaces of the magnetic induction operator
$\L$ \rf{Leig}. It is then evident from \rf{Seq} and \rf{adj} that
for the cosine flows\\
$\bullet$ ${\bf S}_k$ is symmetric in $x_3$ for $k=1,2$ and antisymmetric
in $x_3$ for $k=3$;\\
$\bullet$ ${\bf Z}_l$ is antisymmetric in $x_3$ for $l=1,2$ and symmetric
in $x_3$ for $l=3$.\\
Consequently, by \rf{Dlmk}, $\De^l_{mk}=0$ if all indices $l,m,k$ do not
exceed 2, or precisely two of them are equal to 3.

Thus, eddy diffusivity tensor $\D$ for a cosine flow involves 5 pairs of non-zero
entries of opposite signs, and by virtue of \rf{dei}
$$\lambda_2({\bf q})=-\eta+\De^2_{31}\cos^2\theta+\sin^2\theta
\left(\De^3_{12}\cos^2\varphi+(\De^3_{22}+\De^1_{13})\cos\varphi\sin\varphi
+\De^1_{23}\sin^2\varphi\right)$$
for the wave vector \rf{wv}. The minimum eddy diffusivity \rf{miE} is now
$$\eta_{\rm eddy}=\eta-\max\left(\De^2_{31},\ {1\over2}\left(\De^3_{12}+\De^1_{23}
+\sqrt{(\De^3_{12}-\De^1_{23})^2+(\De^3_{22}+\De^1_{13})^2}\right)\right).$$

We have investigated magnetic eddy diffusivity for
a set of cosine flows that satisfy the following conditions:\\
$\bullet$ The horizontal vectors ${\bf a}=(a_1,a_2,0)\ne0$ and ${\bf b}=(b_1,b_2,0)\ne0$
have integer components, $|a_i|\le 3$, $|b_i|\le 3$; $n\le3$ is a positive integer.\\
$\bullet$ $\bf a$ and $\bf b$ are neither orthogonal, nor parallel. (If
${\bf a\cdot b}=0$, the vertical component of \rf{cos} vanishes identically;
such flows are irrelevant for us, since by the Zeldovich \cite{Zel} antidynamo
theorem a planar flow cannot generate magnetic field. For parallel $\bf a$
and $\bf b$, the flow is also planar: ${\bf v}\cdot(b_2,-b_1,0)=0$.)\\
$\bullet$ The largest common divisor of four integers $a_i$ and $b_i$ is 1,
so that the flow does not have a smaller horizontal periodicity cell
aligned with the coordinate axes.\\
$\bullet$ No pair $\bf a,b$ used for construction of a flow in the set is
derived from another such pair by performing the following transformations
or their compositions:\\
$(i)$ swapping $\bf a\leftrightarrow\bf b$ (this does not alter the flow);\\
$(ii)$ inverting the signs $\bf a\to-a$ or $\bf b\to-b$
(the flow is invariant under any of the two reversals
simultaneously with a shift in half a period $\pi/n$ in $x_3$);\\
$(iii)$ swapping the components $a_1\leftrightarrow a_2$ and
$b_1\leftrightarrow b_2$ (flows obtained for two such pairs $\bf a,b$ map into
each other under swapping of the horizontal components of vector fields and
of the horizontal Cartesian axes, $x_1\leftrightarrow x_2$);\\
$(iv)$ for some $i\le2$, inverting the signs of $a_i\to-a_i$ and
$b_i\to-b_i$ (a flow remains invariant under this inversion accompanied by
reflection of the direction $x_i\to-x_i$ and changing the sign
of the $i$th component of vector fields).

\begin{figure}[p]
\centerline{\raisebox{2in}{(a)} \ \includegraphics[width=2.8in,height=2.3in]{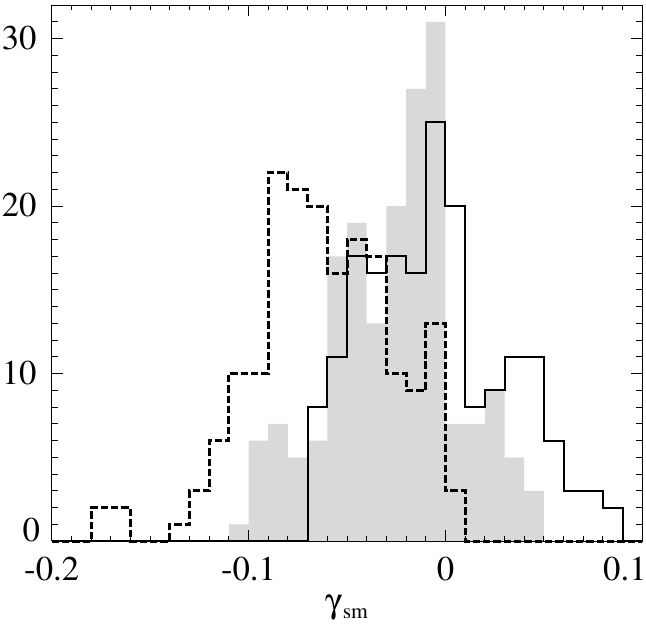}
\hspace{2em}\raisebox{2in}{(b)} \ \includegraphics[width=2.8in,height=2.3in]{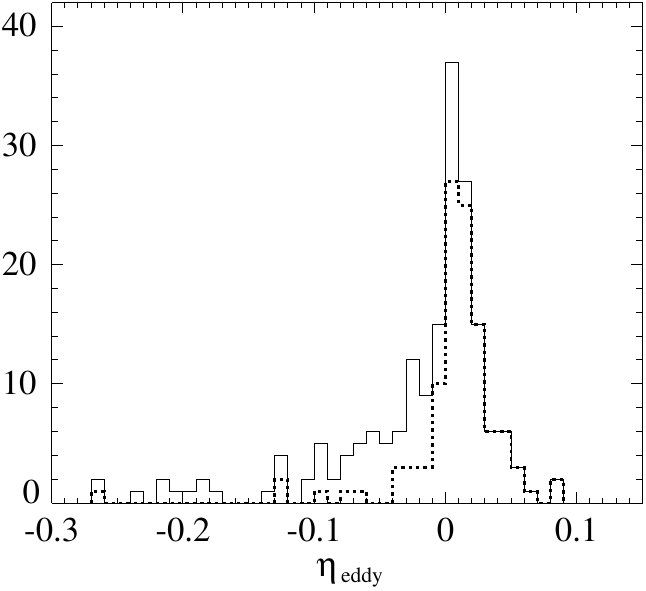}}

~

\centerline{\raisebox{2in}{(c)} \ \includegraphics[width=2.8in,height=2.3in]{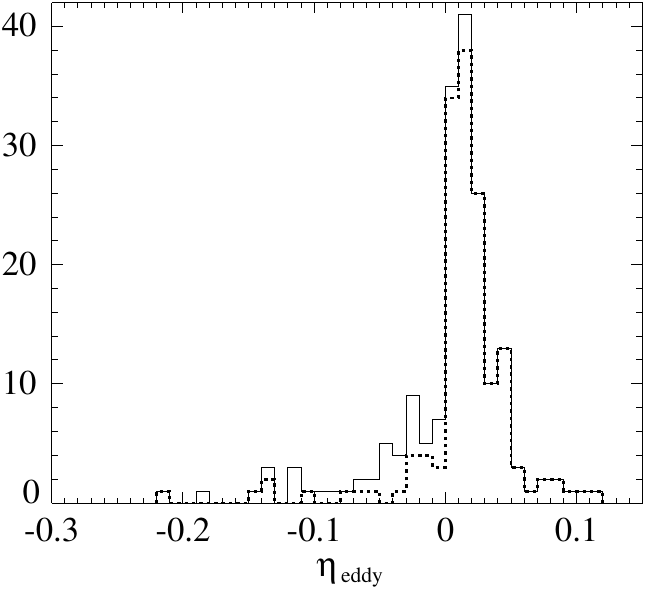}
\hspace{2em}\raisebox{2in}{(d)} \ \includegraphics[width=2.8in,height=2.3in]{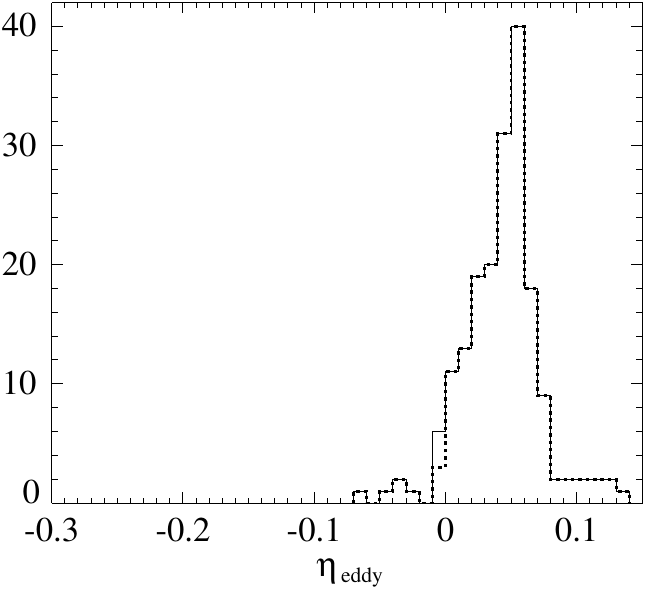}}
\caption{Histogram of dominant fast-time growth rates of small-scale magnetic
modes generated by the 183 primary cosine flows (a) for three values of magnetic
molecular diffusivity: $\eta=0.01$ (black solid line), 0.02 (gray solid line,
the area below the plot is filled in gray), 0.05 (dashed line). Histograms
of the minimum magnetic eddy diffusivity in the 183 primary cosine flows (solid
line) and for those of them which are not small-scale dynamos
(dashed line) for $\eta=0.01$ (b), 0.02 (c) and 0.05 (d).}
\label{hec}

~

\parbox{0.5\textwidth}{
\centerline{\includegraphics[width=3.33in]{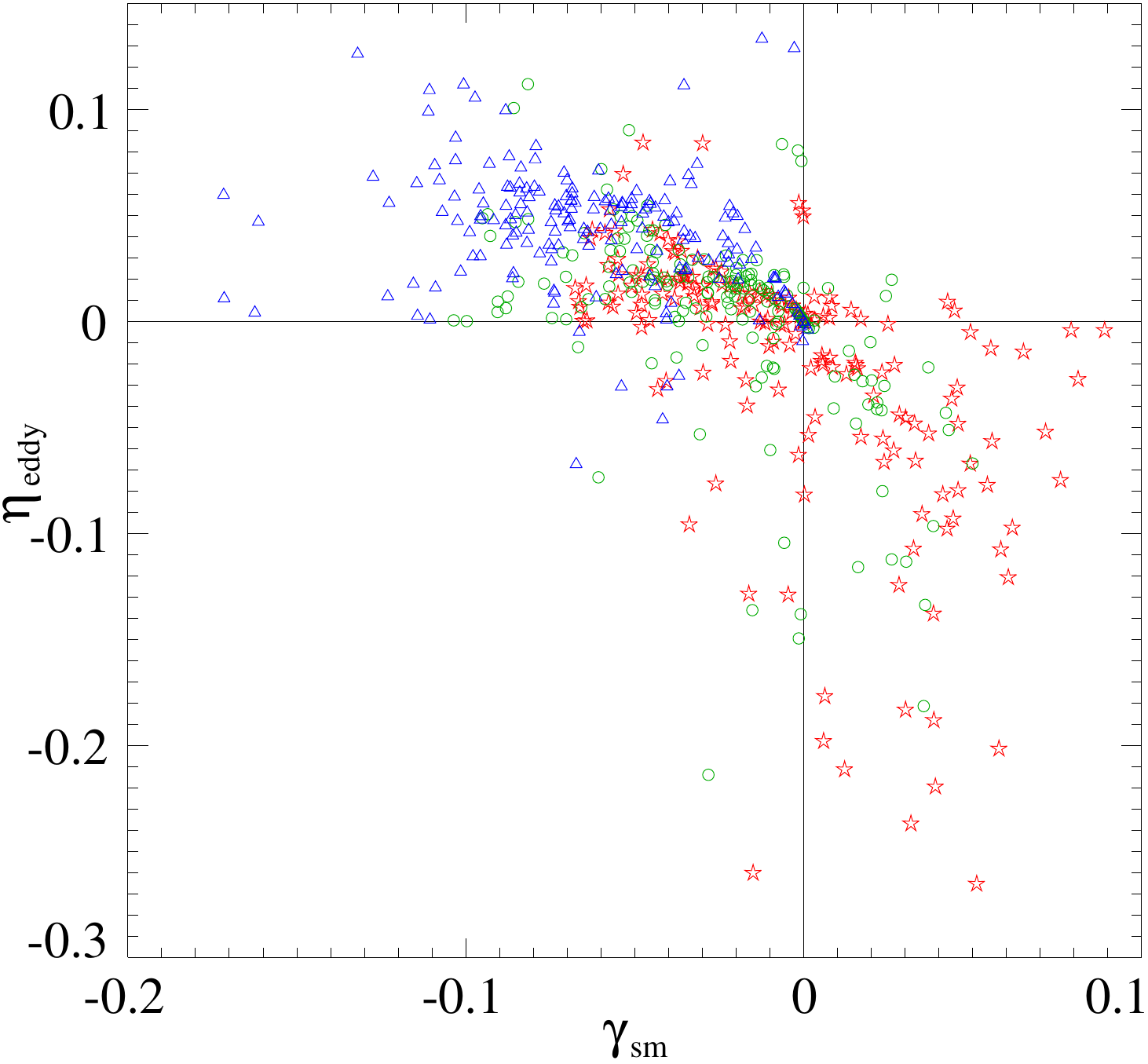}}
\caption{Minimum eddy diffusivity $\eta_{\rm eddy}$ (vertical axis) versus
dominant fast-time growth rates of small-scale modes (horizontal axis) in the primary
cosine flows \rf{cos} for three values of magnetic molecular diffusivity:
$\eta=0.01$ (stars), 0.02 (circles) and 0.05 (triangles).}
\label{ras}}
\parbox{0.02\textwidth}{~}
\parbox{0.48\textwidth}{
\centerline{\includegraphics[width=3.5in]{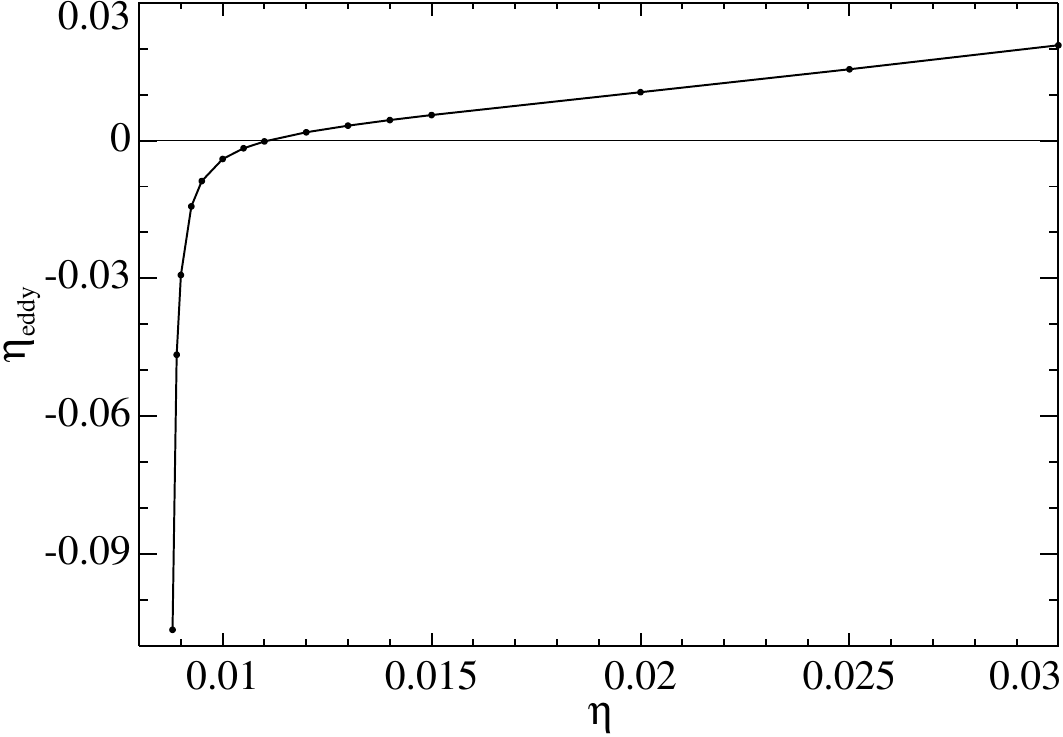}}
\caption{Minimum magnetic eddy diffusivity (vertical axis) in a family V$_1$
sample flow \rf{V1} versus the magnetic eddy diffusivity $\eta$ (horizontal
axis). Dots show the computed values.}
\label{eV1}}\end{figure}

\begin{table}[t]
\caption{Primary cosine flows as small- and large-scale dynamos. Columns 1-4
present numbers of the primary cosine flows \rf{cos} in the specified classes
for two values of magnetic molecular diffusivity.}
\center\begin{tabular}{|c|c|c|c|c|}\hline
&\multicolumn{2}{|c|}{$\eta=0.01$}&\multicolumn{2}{|c|}{$\eta=0.02$}\\\cline{2-5}
&$\eta_{\rm eddy}<0$&$\eta_{\rm eddy}>0$&$\eta_{\rm eddy}<0$&$\eta_{\rm eddy}>0$\\\hline
Small-scale dynamo&61&12&27&\phantom{13}4\\\hline
No small-scale generation&25&85&20&132\\\hline
\end{tabular}\label{tab1}\end{table}

This set of essentially distinct low-wavenumber cosine flows
comprises 183 flows, which we will call {\it primary} cosine flows.
The distribution of the dominant fast-time growth rates of small-scale magnetic
field and of the minimum eddy diffusivity computed for these flows is shown
in Figs.~\ref{hec} and \ref{ras} for three values of molecular diffusivity,
$\eta=0.01$, 0.02 and 0.05 (see also Table~\ref{tab1}). While a significant
proportion of the primary flows is capable of generating small-scale magnetic
field for $\eta\le0.02$, only three instances of them are small-scale dynamos
for $\eta=0.05$. This value is also close to the upper bound of $\eta$,
for which large-scale generation by the cosine flows is possible:
for $\eta=0.05$, only 11 instances of primary cosine flows feature negative
magnetic eddy diffusivity, including the 3 flows that can generate small-scale
field.

The most populated class (consisting of more than a half of the total number
of flows for all considered $\eta$) are the primary cosine flows that are
neither small-, nor large-scale dynamos (quadrant II in \xf{ras}).
By contrast, for $\eta=0.01$ and 0.02, the second largest class are flows that
sustain both small- and large-scale (by the mechanism of negative eddy
diffusivity) generation (quadrant IV in \xf{ras}). Nevertheless, for these
values of molecular diffusivity, there are quite a few primary cosine flows
of our prime interest, that are incapable of small-scale generation but feature
negative eddy diffusivity (quadrant III in \xf{ras}). Finally,
for both $\eta$, the least numerous is the class of flows that can generate
small-scale flows, but give rise only to positive eddy diffusivity (quadrant I
in \xf{ras}). For each of the three classes of flows that sustain
at least one of the two considered types of generation, the cardinality falls
when molecular diffusivity increases from $\eta=0.01$ to 0.02; this agrees with
the ``common sense'' argument that enhancing magnetic molecular diffusivity
hinders the ability to generate magnetic field.

\subsection{Family V$_1$}\label{fV1}

The minimum magnetic eddy diffusivity in a family V$_1$ sample flow \rf{V1}
constructed for $C_1=1$, $C_2=1,\ C_3=-2$ is shown in \xf{eV1}. The functions
$U_i(x_i)$ have been synthesised as Fourier series with zero coefficients
for the wave numbers $|n|>10$ and $n=0$. The coefficients for $0<n\le10$
have been assigned the imaginary values $\I 6^{-n}r_n$, where $r_n$ is
a pseudorandom real number in the interval $[0,1]$, and complex-conjugate
values are employed for $n<0$. The energy spectrum
of the resultant flow decays by 10 orders of magnitude. By construction,
all $U_i(x_i)$ are odd, whereby \rf{V1} is a parity-invariant flow.

We observe a typical behaviour of the minimum eddy diffusivity
\rf{miE}: $\eta_{\rm eddy}>0$ for sufficiently large molecular diffusivity
$\eta$, but $\eta_{\rm eddy}<0$ and thus large-scale generation becomes
possible below a certain $\eta>0$. The minimum eddy diffusivity tends
to $-\infty$ when $\eta$ approaches the critical value
for the onset of the small-scale generation (see section 3.7 of \cite{VZ}).

\section{Concluding remarks}\label{con}

1. We have presented (section \ref{const}) six families of pointwise
non-helical (i.e., satisfying \rf{non}) steady space-periodic three-dimensional
flows of incompressible fluid. One, family P (see section \ref{fP}), consists
of poloidal flows. For determining the potentials $P({\bf x})$ \rf{pot} for certain
such flows, we propose to solve the two-dimensional scalar problems
\rf{red}--\rf{fi} ``of the nonlinear eigenvalue type''. Four families (C, L,
V$_1$ and V$_2$) are
analytically defined. This has provided enough flows to conduct numerical
experiments in magnetic field generation. However, it is desirable to find
methods for constructing pointwise zero-helicity solenoidal flows, whose
streamlines involve knots of a given topology, in the spirit of~\cite{KFDI}.

Furthermore, we have shown (section \ref{hsf}) that flows comprising
families P, C, V$_1$ and V$_2$ have zero helicity spectrum.

2. In application of the multiscale formalism (see \cite{VZ}) to the problem
of kinematic generation of large-scale magnetic field by small-scale flows
in the limit of high scale separation, eigenfunctions and the associated
eigenvalues of the large-scale magnetic induction operator are expanded
in the power series
\rf{ble} in the scale ratio $\varepsilon$. Generically, the $\alpha$-effect
is present, and the leading term in expansion of the eigenvalue is
$\varepsilon\lambda_1$, where $\lambda_1$ is an eigenvalue of the
magnetic $\alpha$-effect operator (see~\rf{aleq}). For an arbitrary unit
(large-scale) wave vector $\bf q$, we have derived the eigenvalues $\lambda_1$
\rf{ei} associated with the harmonic eigenfunctions \rf{mm}, as well as the maximum,
over unit $\bf q$, slow-time growth rate \rf{mgr} sustained by the action
of the $\alpha$-effect. The growth rate is controlled by the symmetric part
of the $\alpha$-effect tensor, $\A$. We have also proven that
the $\alpha$-effect tensor $\A^-$ for the reverse flow $-\bf v(x)$
is obtained from the tensor $\A$ for $\bf v(x)$ by matrix transposition.

When the $\alpha$-effect is absent ($\A=0$), the leading term in the expansion
of the eigenvalue is $\varepsilon^2\lambda_2$, where $\lambda_2$
is an eigenvalue of the magnetic eddy diffusivity operator (see \rf{med}).
We have calculated the eigenvalues $\lambda_2$ \rf{dei} associated with the harmonic
eigenfunctions \rf{mm}. (To the best of our knowledge, the relations \rf{ei} and
\rf{dei} were so far unavailable in the literature; however, see section 9.3 in
\cite{Mb}.)

3. We have computed (section \ref{al}) the maximum (over the direction of the wave
vector $\bf q$) slow-time growth rates Re\,$\lambda_1$ due to the action
of the magnetic $\alpha$-effect in sample flows from families P, V$_1$, V$_2$
and L (defined in sections \ref{fP}, \ref{fV} and \ref{fL}), as well as
(section \ref{ed}) the maximum slow-time growth rates Re\,$\lambda_2$ sustained
by the magnetic eddy diffusivity in sample flows from families L, C (defined
in section \ref{fC}) and V$_1$. (Family C cosine flows are parity-invariant and
thus lack the $\alpha$-effect; by contrast, no family V$_2$ flows are
parity-invariant and might be used to investigate magnetic eddy diffusivity.)

In all considered families we have encountered flows that, for sufficiently
small magnetic molecular diffusivities, $\eta$, generate large-scale magnetic
field by employing the respective mechanism.
In addition, in all the families we have found sample flows
that generate small-scale field. Thus, we have demonstrated that zero kinetic
helicity density does not rule out generation of large-scale magnetic
field by the mechanisms of the $\alpha$-effect or negative magnetic eddy
diffusivity, and of small-scale field. This can be explained heuristically
as follows: Various topological properties of knottedness of vorticity lines
are controlled by the independent quantities $\cal W$, $\cal T$ and $\cal N$,
whose sum $n$ enters the kinetic helicity \rf{WTN} as a factor. Zero helicity
does not
require vanishing of any of these quantities, and non-zero values indicate that
the lines possess non-trivial respective knottedness properties. This implies
an intricate topological structure of the non-helical flow that,
empirically, is likely to give rise to generation of magnetic field.

Actually, from this perspective the flow helicity rather than the kinetic
helicity (which is the vorticity helicity) seems to be more appropriate
for characterising the flow complexity significant for magnetic field
generation. Indeed, in the limit of small local magnetic Reynolds numbers,
in view of the relation
$${\rm tr}\,\A=-\eta^{-1}\LA{\bf v}\cdot(-\nabla\times \nabla^{-2}{\bf v})\RA
+{\rm O}(\eta^{-2})$$
for the tensor \rf{as} (see also chapters 10 and 11 of \cite{VZ}),
the $\alpha$-effect is directly linked with the flow helicity equal to
the spatial mean in the r.h.s. Furthermore, the same expression (up to
the O($\eta^{-2}$) term) was found for the trace of the $\alpha$-effect tensor
in the mean-field electrodynamics (see \cite{RBr} and references therein)
in the low-conductivity limit, and the mean flow helicity was claimed to play
a fundamental role for the dynamo action of the $\alpha$-effect especially
for steady flow (cf.~expression \rf{Anew}
for the $\alpha$-effect tensor for time-periodic flow). Consequently,
a numerical investigation analogous to the present one is desirable, in which
the $\alpha$-effect and eddy diffusivity tensors, \rf{Adef} and \rf{Ddef},
respectively, will be studied numerically for a set of steady three-dimensional
space-periodic solenoidal flows, whose flow helicity density vanishes pointwise.

4. Hydrodynamic helicity exists in two loosely related incarnations:
kinetic helicity with the density ${\bf v}\cdot\nabla\times{\bf v}$, which
is mainly of our concern in the present work, and the helicity spectrum \rf{hm},
which can be regarded as a kind of helicity density in the Fourier space. It was
shown in the theory of mean-field electrodynamics that a non-zero helicity
spectrum is necessary for the action of the $\alpha$-effect under certain
conditions \cite{M70,MP}. Asymptotic multiscale methods (see section \ref{hs})
yield the same conclusion and the same expression for the $\alpha$-effect
tensor in the limit of small R$_m^{\rm loc}$.

However, our numerical results demonstrate that for finite (non-vanishing)
R$_m^{\rm loc}$ a non-zero helicity spectrum is required neither for the action
of small-scale dynamo, nor for generation of large-scale field. Although flows
comprising families P, C, V$_1$ and V$_2$ have an identically vanishing
helicity spectrum
(see section \ref{hsf}), computations attest that this does not result
in vanishing of the $\alpha$-effect tensor and does not prevent
the $\alpha$-effect from generation. We also observe that the helicity spectrum
is zero for all parity-invariant flows, but nevertheless they can sustain
negative eddy diffusivity.

Thus, all flows that we consider numerically feature a zero kinetic helicity
density and zero helicity spectrum, except for two instances of family L flows
discussed in section \ref{alL}, whose helicity spectrum (but not kinetic
helicity density) are non-zero. The slow-time growth rates of large-scale
magnetic fields that they generate are roughly an order of magnitude
higher than the slow-time growth rates due to the action of the $\alpha$-effect
in other flows that we have considered, but it is unclear whether this is just
a coincidence (the influence of the singularities in the graphs \xf{gLa}),
or indeed generation becomes more vigorous under the influence
of a non-zero helicity spectrum. Following the anonymous First Referee, we can
formulate a conclusion as a relaxed modification of the title of \cite{GFP}:
helicity is unnecessary for dynamo action, but it may help.

Kinetic helicity and helicity spectrum turn out to be unsatisfactory measures
of the flow chirality deemed necessary for generation, except in certain limit
cases. By analogy, we doubt that any helicity-type
integral quantity can serve for predicting the ability of a steady flow
to generate a small- or large-scale magnetic field, except in specific
limits --- as the helicity spectrum becomes necessary for persistence
of the $\alpha$-effect when R$_m^{\rm loc}\to0$. Any such criterion is
unlikely to be useful, just predicting that under certain conditions
the first term in the respective expansion of the eigenvalue of the induction
operator vanishes --- in which case generation may still be made possible
by virtue of subleading terms (e.g., in the absence of the $\alpha$-effect
generation may still be powered by negative eddy diffusivity).

Magnetic analogues of the kinetic helicity and helicity spectrum do show up
in the mathematical multiscale kinematic
dynamo theory. The $\alpha$-effect tensor \rf{Adef} can be decomposed as
\BE\A_k=\I\sum_{{\bf n}\ne0}|{\bf n}|^{-1}H_{\bf n}^{{\bf v,S}_k},\EE{hvS}
where
$$H_{\bf n}^{{\bf v,S}_k}=\overline{\hat{\bf v}}_{\bf n}\cdot
(\I{\bf n}\times{\hat{\bf S}}_{k,\bf n})$$
can be interpreted as the cross-helicity spectrum of the flow $\bf v$ and
neutral magnetic mode ${\bf S}_k$ (their Fourier coefficients are denoted
by ${\hat{\bf v}}_{\bf n}$ and ${\hat{\bf S}}_{k,\bf n}$, respectively).
In view of the relations \rf{cS} (to the best of our knowledge
discovered in \cite{VZ}), helicities of the currents $\nabla\times{\bf S}_k$
associated with small-scale neutral modes of the magnetic induction operator
$\L$ are also more appropriate than the kinetic helicity or helicity
spectrum for characterising the magnetic $\alpha$-effect: in the limit of high
spatial scale separation, \rf{hvS} and \rf{cS} hold true for any
local magnetic Reynolds number and not just for R$_m^{\rm loc}\to0$.
However, such identities just manifest the mathematical fact that the mean
vector product of two solenoidal space-periodic fields is a linear
functional of their cross-helicity spectrum; we doubt that this reveals any
physical significance of their helicity spectra. The $\alpha$-effect tensor
does vanish when all $H_{\bf n}^{{\bf v,S}_k}=0$, but to check this we must
compute the neutral modes ${\bf S}_k$ --- and afterwards it is straightforward
to directly compute the tensor \rf{Adef}.

5. The maximum slow-time growth rate \rf{mgr} due to the action of the magnetic
$\alpha$-effect is non-negative; it vanishes if and only if the intermediate
eigenvalue $\alpha_2$ of the symmetrised $\alpha$-effect tensor $\sA$ is zero.
Figure \ref{rva} demonstrates that on decreasing $\eta$ there can be quite a few points,
where $\alpha_2=0$. Thus, the maximum growth rate \rf{mgr} may be expected
to exhibit an intermittent behaviour analogous to the one shown in \xf{rva},
with strictly positive values separated by zero for infinitely many
$\eta$'s accumulating at $\eta=0$.

6. It is known that magnetic eddy diffusivity can become infinitely negative
when magnetic molecular diffusivity approaches the critical value
$\eta\to\eta_{\rm cr}$ for the onset of the small-scale dynamo action,
and we have encountered such behaviour in the present work. When considering
generation of large-scale magnetic field by the magnetic $\alpha$-effect, we
have come across a similar infinite increase of the maximum slow-time growth
rate $\gamma_\alpha$ \rf{mgr} in the limit $\eta\to\eta_{\rm cr}$ (see
\xf{gLa}b). The two phenomena have the same nature. We have explained
such a singular behaviour of $\gamma_\alpha$ by considering
expansions of the fluctuating part ${\bf S}_k$ of neutral modes in the basis
of eigenfunctions of the small-scale magnetic induction operator $\L$.

7. We have encountered an instance of two disjoint intervals of magnetic molecular
diffusivity where small-scale generation is absent, but large-scale magnetic
field is generated by the mechanism of negative magnetic eddy diffusivity.
The window between these two intervals, where no large-scale generation
happens, may be regarded as a large-scale dynamo counterpart to the window
of quiescence of the small-scale kinematic dynamo action by the 1:1:1 ABC-flow
discovered in \cite{ArK}.

8. Are our dynamos fast or slow? Families V$_1$ \rf{V1} and V$_2$ \rf{V2} flows
are integrable (to see this, introduce a new time $\tau$ satisfying
$\d\tau/\d t=\dot U_1\dot U_2\dot U_3$ for a V$_1$ flow or
$\d\tau/\d t=U_1U_2U_3$ for a V$_2$ flow). Therefore, they cannot act as
fast dynamos, because this requires chaotic behaviour of fluid particle
trajectories and positiveness of topological entropy of the flow \cite{KY}
(see also \cite{V89,So}). Equally, trajectories of a cosine flow \rf{cos}
and of a family P flow \rf{pol} for the potential \rf{pot} lie on vertical
cylindrical surfaces, whose intersection with a horizontal plane can be found
by considering the ratio of the two differential equations for horizontal
coordinates of a fluid particle. This rules out chaos. For other family P
flows and for family L flows a more detailed study of integrability and
chaotic properties is required.

We can examine this question from a different perspective. Computations suggest
that for the molecular diffusivity $\eta$ tending to zero, the small-scale
neutral modes ${\bf S}_k+{\bf e}_l$ (and their reverse-flow counterparts
${\bf S}^-_l+{\bf e}_l$) grow in amplitude. In view of \rf{Adef},
the magnitude of the $\alpha$-effect tensor is an outcome of the competition
of this growth and the decay of the energy spectrum of the smooth flow~$\bf v$.
Also, we observe in ${\bf S}_k$ a shift of the maximum of the energy spectrum
towards large wave vectors; if it occurs simultaneously with a similar shift
in ${\bf S}^-_l$, then by \rf{Dlmk} and \rf{Zl} the magnitude of the eddy
diffusivity tensor may increase in this limit (although the definition \rf{Ddef}
does not make this obvious). Consequently, a fast (in the respective slow time)
dynamo is apparently not ruled out for both mechanisms of large-scale
generation. However, a version of the argument presented in section
\ref{alL} is applicable: the growth rates measured in the fast time are
approximated by the leading terms in \rf{lex}, $\varepsilon$Re$\,\lambda_1$
for the $\alpha$-effect dynamos and $\varepsilon^2$Re$\,\lambda_2$
for the eddy diffusivity dynamos, for sufficiently small $\varepsilon$ only.
Bounds for such relevant $\varepsilon$ on varying $\eta$ are
an open question; the growth rates observed in the fast time for these
$\varepsilon$ are likely to tend to zero even if $\lambda_1$ or
$\lambda_2$, respectively, increase when $\eta\to0$.

9. How will the subsequent nonlinear evolution modify our findings about
kinematic generation? This is affected by many factors.

Will kinetic helicity density remain zero?
Even in the purely hydrodynamic (in the absence of magnetic field) nonlinear
evolution of ideal fluid flow governed by the Euler equation, nothing prevents
the flow from losing this property: it is only guaranteed, as we have mentioned
in the introduction, that the mean kinetic helicity stored in any volume, whose
boundary is everywhere tangent to the vorticity (including the periodicity
cell which can be regarded as having no boundary), does not change
in time and will thus remain zero. However, the presence of the Lorentz force
and/or viscosity renders inapplicable this conservation law. The flow is likely
to acquire non-zero helicity density --- thus making inappropriate the
question that we address in the present work.

How long will the magnetic field remain multiscale? This is an intriguing
open question, which can be fully solved by direct numerical simulations
only. We expect the scale separation to persist while
the MHD perturbation remains weakly nonlinear.
A study of amplitude equations for a large-scale weakly nonlinear
perturbation of a small-scale convective dynamo \cite{CZ} has revealed that
the amplitudes governing the perturbation blow up at a finite slow time.
This does not mean that the perturbation itself develops a singularity --- just
the asymptotic expansion ceases to be applicable when the perturbation becomes
too strong --- but perhaps suggests an abrupt transition to a regime without
scale separation. Other reasons may also be responsible for transformation
of a multiscale field into one with a continuous distribution of scales.
A magnetic perturbation composed of a unique mode \rf{bex} is a mathematical
idealisation: In a physical system many such modes corresponding to different
integer (taking into account the periodicity) multiples of the scale
ratio $\varepsilon$ are present due to noise and emerge, because the magnetic
perturbation is coupled with the hydrodynamic one; the same is true for digital
simulations, in which the noise is due to round-off errors. The larger is
the integer factor, the larger is the slow-time growth rate \rf{lex}
of the mode. Thus, competition of large-scale magnetic modes sets in, in which
the modes of smaller spatial periods have higher chances to win. Together with
the inverse cascade, this will apparently destroy the multiscale nature
of the magnetic field.

Will magnetic field generation continue and will magnetic field persist?
Simulations of various nonlinear small-scale convective dynamos attest that
most different scenarios are possible: upon modification by the Lorentz force
due to the growing magnetic field, the flow can settle to a non-generating
hydrodynamic attractor, and in such a ``self-extinguishing dynamo'' magnetic
field decays to zero \cite{DS}; alternatively, the MHD regime can saturate
to a stable MHD steady or periodic state \cite{CGPZ}; or it may consist of
a chaotic sequence of visits to formerly attracting hydrodynamic states, some
of which can generate magnetic field, and other ones cannot, thus exhibiting
intermittent upsurges and decays of magnetic field \cite{CRC}; or the evolution
can consist of a chaotic sequence of visits to unstable but identifiable MHD
states --- a scenario \cite{Pod} for development of magnetic field reversals
is an example; or it can be just an unstructured chaotic trajectory
in the phase space. In any case, when the overcriticality is not too small,
the initial flow is abandoned, rendering inapplicable our analysis.

\section*{Acknowledgements}

RC was partially supported by the project POCI-01-0145-FEDER-006933/SYSTEC
(Research Center for Systems and Technologies, University of Porto) financed
by ERDF (European Regional Development Fund) through COMPETE 2020
(Programa Operacional Competitividade e Internacionaliza\c c\~ao), and by FCT
(Funda\c c\~ao para a Ci\^encia e a Tecnologia, Portugal). The main bulk of
computations has been carried out on the cluster ``Sergey Korolev''
at Samara University. We are grateful to the two anonymous Referees, whose
comments have significantly contributed to the improvement of the paper, and
especially for the suggestion of the Second Referee to include the helicity
spectrum of flows in our investigation.

\end{document}